\documentclass[twocolumn]{aastex62}

\graphicspath{{./}{figures/}}
\def \OIII {[O\,{\sc iii}]}

\newcommand{\OIIIb}{\OIII\,$\lambda$5007}
\newcommand{\AAm}{\text{\AA}}
\newcommand{\javelin}{{\tt JAVELIN}\,}

\def\CIV{C\,{\sc iv}}
\def\MgII{Mg\,{\sc ii}}
\def\FeII{Fe\,{\sc ii}}

\usepackage{color,subfigure}

\usepackage{ulem}
\usepackage{amsmath,bm}
\usepackage{commath}
\usepackage[figuresright]{rotating}
\usepackage{tabularx}
\usepackage[vlines]{tabularht}
\usepackage{multirow}
\usepackage{longtable}
\usepackage{textcomp}
\usepackage{booktabs}
\DeclareUnicodeCharacter{2212}{-}
\usepackage{savesym}
\savesymbol{tablenum}
\usepackage{siunitx}
\restoresymbol{SIX}{tablenum}

\shorttitle{CXO CLQs}
\shortauthors{Yang et al.}

\begin{document}

\title{Probing the Origin of Changing-look Quasar Transitions with Chandra}

\correspondingauthor{Qian Yang}
\email{qian.yang@cfa.harvard.edu} 

\author[0000-0002-6893-3742]{Qian Yang}
\affiliation{Center for Astrophysics $\vert$ Harvard \& Smithsonian, 60 Garden Street, Cambridge, MA 02138, USA}

\author[0000-0002-8179-9445]{Paul J. Green}
\affiliation{Center for Astrophysics $\vert$ Harvard \& Smithsonian, 60 Garden Street, Cambridge, MA 02138, USA}

\author[0000-0003-3422-2202]{Chelsea L. MacLeod}
\affiliation{BlackSky, 1505 Westlake Ave N \#600, Seattle, WA 98109, USA}

\author[0000-0002-7092-0326]{Richard M. Plotkin}
\affiliation{Physics Department, University of Nevada, Reno, 1664 N. Virginia St., Reno NV 89557, USA}
\affiliation{Nevada Center for Astrophysics, University of Nevada, Las Vegas, NV 89154, USA}

\author[0000-0002-6404-9562]{Scott F. Anderson}
\affiliation{Department of Astronomy, University of Washington, Box 351580, Seattle, WA 98195, USA}

\author[0000-0001-6637-5401]{Allyson Bieryla}
\affiliation{Center for Astrophysics $\vert$ Harvard \& Smithsonian, 60 Garden Street, Cambridge, MA 02138, USA}

\author[0000-0002-2115-1137]{Francesca Civano}
\affiliation{NASA Goddard Space Flight Center, Greenbelt, MD 20771, USA}

\author[0000-0002-3719-940X]{Michael Eracleous}
\affiliation{Department of Astronomy \& Astrophysics and Institute for Gravitation and the Cosmos, 525 Davey Laboratory, The Pennsylvania State University, University Park, PA 16802, USA}

\author[0000-0002-3168-0139]{Matthew Graham}
  \affiliation{Cahill Center for Astronomy and Astrophysics, California Institute of Technology, 1216 E. California Boulevard, Pasadena, CA 91125, USA}
  
\author[0000-0001-8665-5523]{John J. Ruan}
\affiliation{Department of Physics and Astronomy, Bishop's University, 2600 College St., Sherbrooke, QC J1M 1Z7, Canada}

\author[0000-0001-8557-2822]{Jessie Runnoe}
\affiliation{Department of Physics and Astronomy, Vanderbilt University, Nashville, TN 37235, USA}

\author[0000-0002-7791-3671]{Xiurui Zhao}
\affiliation{Center for Astrophysics $\vert$ Harvard \& Smithsonian, 60 Garden Street, Cambridge, MA 02138, USA}

\begin{abstract}
Extremely variable quasars can also show strong changes in broad-line emission strength and are known as changing-look quasars (CLQs). To study the CLQ transition mechanism, we present a pilot sample of CLQs with X-ray observations in both the bright and faint states. From a sample of quasars with bright-state archival SDSS spectra and ({\it Chandra} or {\it XMM-Newton}) X-ray data, we identified five new CLQs via optical spectroscopic follow-up, and then obtained new target-of-opportunity X-ray observations with {\it Chandra}.  No strong absorption is detected in either the bright- or the faint-state X-ray spectra.
The intrinsic X-ray flux generally changes along with the optical variability, and the X-ray power-law slope becomes harder in the faint state. Large amplitude mid-infrared variability is detected in all five CLQs, and the MIR variability echoes the variability in the optical with a time lag expected from the light-crossing time of the dusty torus for CLQs with robust lag measurements. The changing-obscuration model is not consistent with the observed X-ray spectra and spectral energy distribution changes seen in these CLQs. It is highly likely that the observed changes are due to the changing accretion rate of the supermassive black hole, so the multiwavelength emission varies accordingly, with promising analogies to the accretion states of X-ray binaries. 

\end{abstract}


\section{Introduction} \label{sec:introduction}

Active galactic nuclei (AGN) are powered by supermassive black holes (SMBHs) accreting gas, producing radiation that spans the entire electromagnetic spectrum \citep[e.g.,][]{Elvis1994, Lusso2012}.  The X-ray emission comes from a hot corona \citep[][]{Haardt1993} and possibly a jet \citep{Narayan1994, Blandford1999}. 
In the UV/optical, there is continuum emission from the accretion disk \citep[e.g.,][]{Shields1978}, as well as broad and narrow emission lines from the broad-line region (BLR) and narrow-line region (NLR), respectively \citep[][]{Davidson1979,Boroson1992}.  
Dust in the torus is heated by the emission from the accretion disk and reemitted as thermal radiation in the infrared \citep{Pier1993}.

Observationally, AGN may be classified based on their multiwavelength characteristics.
In the UV/optical, there are type 1 AGN with both broad ($ \gtrsim 1000 ~{\rm km~s^{-1}}$) and narrow emission lines ($\lesssim 1000 ~{\rm km~s^{-1}}$); there are also type 2 AGN with only narrow emission lines \citep[][and references therein]{Netzer2015}. 
In the X-ray regime, there are unobscured AGN with $N_{\rm H} < 10^{22}~{\rm cm^{-2}}$ and obscured AGN with $N_{\rm H} \geq 10^{22}~{\rm cm^{-2}}$ \citep[][and references therein]{Ricci2022}. 
Most type 1 AGN are X-ray unobscured, and most type 2 AGN are X-ray obscured \citep[e.g.,][]{Koss2017, Ricci2017, Oh2022}. The canonical AGN unified model interprets the different classes of AGN as the same class of object viewed at different angles \citep{Antonucci1993, Urry1995}. In this model, AGN observed face-on, revealing unobscured emission from both the BLR and the NLR, are type 1. When viewed edge-on, the central region and broad emission line region are obscured by the dusty torus, so we observe type 2 AGN with only narrow emission lines and high intrinsic columns of dust and gas.

However, there are some rare objects that suggest additional or complementary possibilities to this geometric model.
Some type 2 AGN show no broad emission lines but also no detectable obscuration even in X-rays \citep{LaMassa2014, Panessa2009}.  Some of these ``naked" type 2 AGN, selected for lack of apparent broad emission lines but with AGN-like optical variability, do show faint broad emission line components in higher signal-to-noise ratio optical spectra \citep{Barth2014, Lopez-Navas2022}.  

Direct evidence that broad lines can strongly dim or even disappear in AGN, leaving only narrow lines come from ``changing-look AGN" \citep[CL AGN; e.g.,][]{Denney2014, Shappee2014, McElroy2016, Shapovalova2019, Trakhtenbrot2019}.
Remarkably, even the higher-luminosity quasars can change between different types on timescales of a year or less.
Such objects exhibit large amplitude changes in luminosity, accompanied by the dramatic emergence or near disappearance of broad emission-line (BEL) components, dubbed ``changing-look quasars"
\citep[CLQs; e.g.,][]{LaMassa2015, Ruan2016, Runnoe2016, MacLeod2016, Gezari2017, Yang2018}.  
The ``Changing-look" phenomena were also found in $z>2$ quasars with dramatic changes in their \CIV\ emission lines \citep{Ross2020}.
Repeat spectroscopy for large quasar samples continues to uncover this rare and remarkable phenomenon in regimes of luminosity and redshift that now overlap the large cosmological samples of quasars of the Sloan Digital Sky Survey \citep[SDSS; ][]{Schneider2010, Lyke2020, Green2022, Almeida2023}. 

The discovery of CLQs profoundly affects many areas of astrophysics, including our understanding of quasar duty cycles \citep[e.g.,][]{Martini2003} and galaxy evolution \citep{Kormendy2013}, through distinct feedback processes in states of high and low accretion rate \citep[e.g.,][]{Fabian2012}.  
Rapid swings in SMBH accretion rate may also explain the complex AGN/star-formation connection \citep[e.g.,][]{Hickox2014}, and help reconcile cosmological simulations with the observed prevalence of an ionized intergalactic medium or ``AGN proximity zone fossils" \citep{Oppenheimer2018}. These intriguing prospects explain why CLQs have been ardently sought in the last few years, resulting in at least 80 now known \citep[e.g.,][]{LaMassa2015, Ruan2016, MacLeod2016, Yang2018, Frederick2019, MacLeod2019, Sheng2020, Hon2022, Green2022, Zeltyn2022}.

Several explanations for the strong and relatively rapid luminosity and BEL changes have been proposed.
The first direct explanation is the changing obscuration due to moving clouds crossing over our line of sight. The term ``changing-look" was originally used to describe objects whose X-ray spectra showed the absorption changing between Compton thick and Compton thin.  Absorption changes are a widely accepted explanation for X-ray changing-look events \citep[e.g.,][]{Matt2003, Bianchi2005, Piconcelli2007, Ballo2008, Marchese2012, Ricci2016}. Clouds in the BLR or torus may plausibly eclipse the much smaller X-ray emission region \citep{Maiolino2010}. 
However, the crossing time for intervening clouds orbiting outside the BLR is too long \citep{LaMassa2015} compared to observed CLQ transition timescales.
A tidal distruption event (TDE), in which a star is disrupted by and accretes onto a SMBH, is also a  mechanism for strong, rapid variability.  However, TDEs can only explain a few turn-on CLQ cases \citep[e.g.,][]{Eracleous1995, Merloni2015, Blanchard2017}, and TDE flares are inconsistent with most CLQ light-curve shapes and bright state durations \citep{MacLeod2016, Yang2018, Green2022}. A nuclear supernova (SNe) explosion could also cause a luminous nuclear flare, but the timescale and MIR color variability of SNe are inconsistent with CLQs \citep{Yang2019}. Large accretion rate changes seem the most likely explanation as suggested by recent work \citep{LaMassa2015, Runnoe2016, Yang2018, MacLeod2019, Green2022}. 
However, for major accretion rate changes, the radial inflow timescales \citep[$\sim10^4$ yrs;][]{Krolik1999} are inconsistent with the observed transition timescale of CLQs - from a few months to decades 
(e.g., \citealt{Yang2018}, this work).
A combination of factors may be necessary to explain some CLQ behavior, e.g., rapid changes in the UV ionizing continuum mediate both the illumination of the BLR and the sublimation or re-formation of dust in the inner torus, just outside the BLR \citep{Zeltyn2022}.

Strong accretion state transitions are also seen in X-ray observations of the 
Galactic X-ray binaries \citep[XRBs; e.g.,][]{Maccarone2003, Debnath2010, Kara2019, Wang2022},
suggesting analogies between AGN and XRB accretion \citep[e.g.,][]{Ruan2019, Jin2021}. 
While typical BH accretion regions cannot yet be spatially
resolved, temporal changes in XRB spectral states have gone a
long way toward uncovering the accretion physics in XRBs, and suggest
powerful theoretical and observational analogies to AGN and quasars
\citep{Narayan2005, Merloni2003, McHardy2010, Ruan2019}. However, mass scaling from XRBs to SMBHs 
\citep[e.g.,][]{Sobolewska2011, Schawinski2010} suggests  timescales 
($\sim$$10^{\,4-7}$~yr) for quasar accretion state transitions. Clearly, CLQ transition timescales do not scale as expected from XRBs \citep{Stern2018}. Thus, large spectral state changes in quasars challenge and invigorate debates both about accretion theory and the nature of historical quasar classes \citep[i.e., Type~1 vs Type 2;][]{Elitzur2014}.   Some recent studies suggest that the timescale problem can be qualitatively alleviated when taking the large-scale magnetic field into account \citep{Dexter2019, Pan2021} or considering accretion disk instabilities \citep[e.g., the radiation pressure instability;][]{Sniegowska2020}.

Comparison of multiwavelength observations before and after a transition is needed
to test different models of CLQs. With X-ray spectra in both states, we can detect changes in obscuration or in X-ray spectral power-law slopes.
X-ray spectral shape variations were observed in a handful of CL AGN at lower luminosities (Seyfert galaxies), including Mrk 1018 \citep{Husemann2016}, Mrk 590 \citep{Denney2014}, NGC 1566 \citep{Parker2019}, and 1ES 1927+65 \citep{Trakhtenbrot2019}. 
Hardening of the X-ray-to-UV spectrum of the fading CL AGN Mrk 1018 suggests accretion state transitions similar to XRBs \citep{Noda2018}.  Mrk 1018 also shows evidence that both X-ray photon index and $\alpha_{\rm OX}$, similar to some XRBs, exhibit V-shaped correlations (negative/positive correlation below/above a critical value) in a single AGN \citep{Lyu2021}.
 In comparison to quasars, however, these AGN tend to have $\sim10-100$ times lower SMBH masses, high Eddington ratios, and unusual, X-ray bright spectral energy distributions. 

The first CLQ, SDSS J015957.64+003310.5 (hereafter, J0159), discovered by \citet{LaMassa2015} is an X-ray selected AGN, and there are serendipitous X-ray observations: {\it XMM-Newton} observation in its bright state and {\it Chandra} observation in its faint state. \citet{LaMassa2015} analyzed its X-ray spectra and found its X-ray flux dropped by a factor of 7 with no change in absorption.
\citet{Ai2020} analyzed the X-ray data of a CLQ, SDSS J155258+273728 (hereafter, J1552) discovered by \citet{Yang2018}, and found the absorption in X-ray is moderate and stable, and the X-ray spectrum becomes harder in the bright state, with an Eddington ratio being lower than a few percent in both states. They found that the CL AGN and CLQ with X-ray observations in both states follow a ``V"-shaped correlation: above a critical turnover luminosity the X-ray spectrum is softer when brighter, and below the critical luminosity the trend is reversed as harder when brighter.  \citet{Ruan2019} obtained {\it Chandra} X-ray observations of five CLQs,
and found consistency with accretion state transitions in prototypical X-ray binary outbursts, as well as a possible V-shaped correlation for the UV-to-X-ray spectral index with an inversion correlation at a critical Eddington ratio of $\sim 10^{-2}$. However, for those five CLQs only 
ROSAT All-Sky Survey measurements are available for the bright state, yielding
upper limits on the bright-state X-ray luminosity (and thus lower limits on $\alpha_{\rm OX}$).
\citet{Jin2021} performed {\it Chandra} observations for ten CLQs and further investigated the similarities between AGN and XRBs. However, among those ten CLQs, only two (J2252 and J2333) had archival {\it XMM-Newton} X-ray detections in their former bright states; there are again only upper limits on the bright-state X-ray luminosity (and thus lower limits on $\alpha_{\rm OX}$) for the remaining eight CLQs, preventing much analysis of changes in obscuration or X-ray spectral power-law slopes for CLQs.

To test the different models of CLQs, probe the structure of AGN, and the accretion processes of SMBHs, we pursue a study of CLQs with X-ray observations in both states.  
In this work, we present a pilot sample of five new CLQs with X-ray observations in both states. 
Our sample doubles the size of optical spectroscopic confirmed CLQs. Here for clarity, CLQs are CL AGN with $L_{\rm bol} \gtrsim 10^{45} {\rm erg}~{\rm s}^{-1}$ in the bright state.
with X-ray observations in both states.
The five new CLQs are SDSS J020621.67-060952.7 (hereafter, J0206) at $z = 0.413$, SDSS J022429.10-091851.7 (hereafter J0224) at $z = 0.357$, SDSS J082905.98+420204.3 (hearafter J0829) at $z=0.638$, SDSS J122638.66-001114.0 (hearafter J1226) at $z=0.642$, and SDSS J133806.59-012412.8 (hearafter J1338) at $z=0.452$. All of these quasars have been observed to dim, due to our selection criteria, as described below.

This paper is organized as follows. We present target selection, data, and reduction in Section 2. In Section 3, we describe the data analysis methods, including optical spectral fitting, X-ray spectral analysis, and SED fitting. Results and discussions are in Section 4 and 5, respectively. We summarize the paper in Section 6. Throughout this paper, we adopt a standard $\Lambda$CDM cosmology with $\Omega_{\Lambda}=0.7$, $\Omega_m=0.3$, and $H_0=70$ km s$^{-1}$ Mpc$^{-1}$. 

\section{Selection, Data, and Reduction} \label{sec:data}
\subsection{CLQ Target Selection}
We selected candidates from the SDSS quasar catalogs \citep{Schneider2010, Paris2018}, applying a redshift limit of $z < 0.9$ to keep H$\beta$ within the SDSS spectral range. 
Radio-detected objects were excluded using the FIRST survey \citep[][with a detection limit of 1 mJy]{Becker1995}, since jet-dominated emission in the bright state and its variability would complicate our interpretations. 
To study CLQs in both states, we required existing archival {\it XMM-Newton}  or \textit{Chandra} observations close in time to the SDSS spectroscopic observations. 
For these X-ray observed quasars, we compiled optical light curves from multiple imaging surveys (see \textsection \ref{sec:optical_imaging}), and selected CLQ candidates with a large-amplitude optical variability, specifically $\lvert \Delta g \rvert >$ 1 mag and $\lvert \Delta r \rvert >$ 0.5 mag (as described in \citealt{MacLeod2019}), comparing the latest available epoch (at the time before optical spectroscopic follow-up) to the bright state.

We obtained new optical follow-up spectra using 6-8 m telescopes (\textsection \ref{sec:optical_spec}) as the CLQ candidates dramatically faded in the optical. Once we confirmed with a follow-up optical spectrum a 3$\sigma$ fading of the broad H$\beta$ flux, we triggered a \textit{Chandra} X-ray Target of Opportunity (ToO) observation (\textsection \ref{sec:Xray_obs}), as well as radio observations with the  Karl G. Jansky Very Large Array (VLA) (\textsection \ref{sec:radio_obs}). We summarize the basic properties of the five CLQs observed in Table \ref{tab:Objects}. 

\begin{deluxetable*}{lccccl}
\tablecaption{Changing-look Quasars \label{tab:Objects}}
\tablewidth{1pt}
\tablehead{
\colhead{Full Name} &
\colhead{Short Name} &
\colhead{RA} &
\colhead{Decl.} &
\colhead{Redshift}
}
\startdata
SDSS J020621.67$-$060952.7 & J0206 & 02:06:21.67 & $-$06:09:52.70 & 0.413 \\ 
SDSS J022429.10$-$091851.7 & J0224 & 02:24:29.10 & $-$09:18:51.70 & 0.357 \\
SDSS J082905.98+420204.3 & J0829 & 08:29:05.98 & +42:02:04.30 & 0.638 \\ 
SDSS J122638.66$-$001114.0 & J1226 & 12:26:38.66 & $-$00:11:14.06 & 0.642 \\ 
SDSS J133806.59$-$012412.8 & J1338 & 13:38:06.59 & $-$01:24:12.84 & 0.452 \\ 
\enddata
\end{deluxetable*}

\subsection{Optical Imaging Data} \label{sec:optical_imaging}
We used multiple publicly available optical imaging surveys, including the SDSS imaging survey \citep[in $ugriz$ bands;][]{Abazajian2009}, the Pan-STARRS1 \citep[PS1 in $grizy$ bands;][]{Chambers2016}, the Palomar Transient Factory survey \citep[PTF in $g$ and $R$ bands;][]{Law2009}, and the Zwicky Transient Facility survey \citep[ZTF in $gri$ bands;][]{Bellm2019}. We compiled $g$- and $r$-band light curves from these surveys, using the point-spread function (PSF) magnitudes. For objects without recent public ZTF photometric data, we performed imaging observations in $g$ and $r$ bands using the 1.2 meter telescope at the Fred Lawrence Whipple Observatory (FLWO) on Mount Hopkins in Arizona. 

To calibrate the optical data from different surveys onto the same flux scale for light-curve plotting, we applied corrections to account for different filter curves. We convolved each quasar spectrum with the PS1, PTF, ZTF, and FLWO filter curves to obtain synthetic magnitudes, and derived the magnitude offsets, typically $\sim 0.02$ mag in the $g$ band and 0.09 mag in the $r$ band between similar filters from different surveys.  We used these offsets to calibrate all optical magnitudes to SDSS filter system magnitudes for the purposes of light-curve plotting. 

We also used the Catalina Real-time Transient Survey \citep[CRTS;][]{Drake2009}. The CRTS photometric data are unfiltered, so we applied a constant offset to match the contemporaneous $r$-band light curve described above.  We describe photometric data we used for spectral energy distributions (SEDs) in \textsection \ref{sec:SED_fit} and summarize the photometric data in Table \ref{tab:SED_data}.

We corrected the Galactic extinction using the dust reddening map of \citet[][SFD]{Schlegel1998} and the reddening law in \citet{Fitzpatrick1999}. The typical $E(B-V)_{\rm SFD}$ for the five CLQs is 0.02 mag. 
We computed the extinction coefficients for the surveys we used in this work (see more details in Appendix \textsection \ref{sec:ext}). We adopt a $R_V=3.1$, employing the traditional value of $R_V$ in the diffuse interstellar medium \citep{Cardelli1989}.

\begin{figure*}[!ht]
\centering
\includegraphics[width=0.95\textwidth]{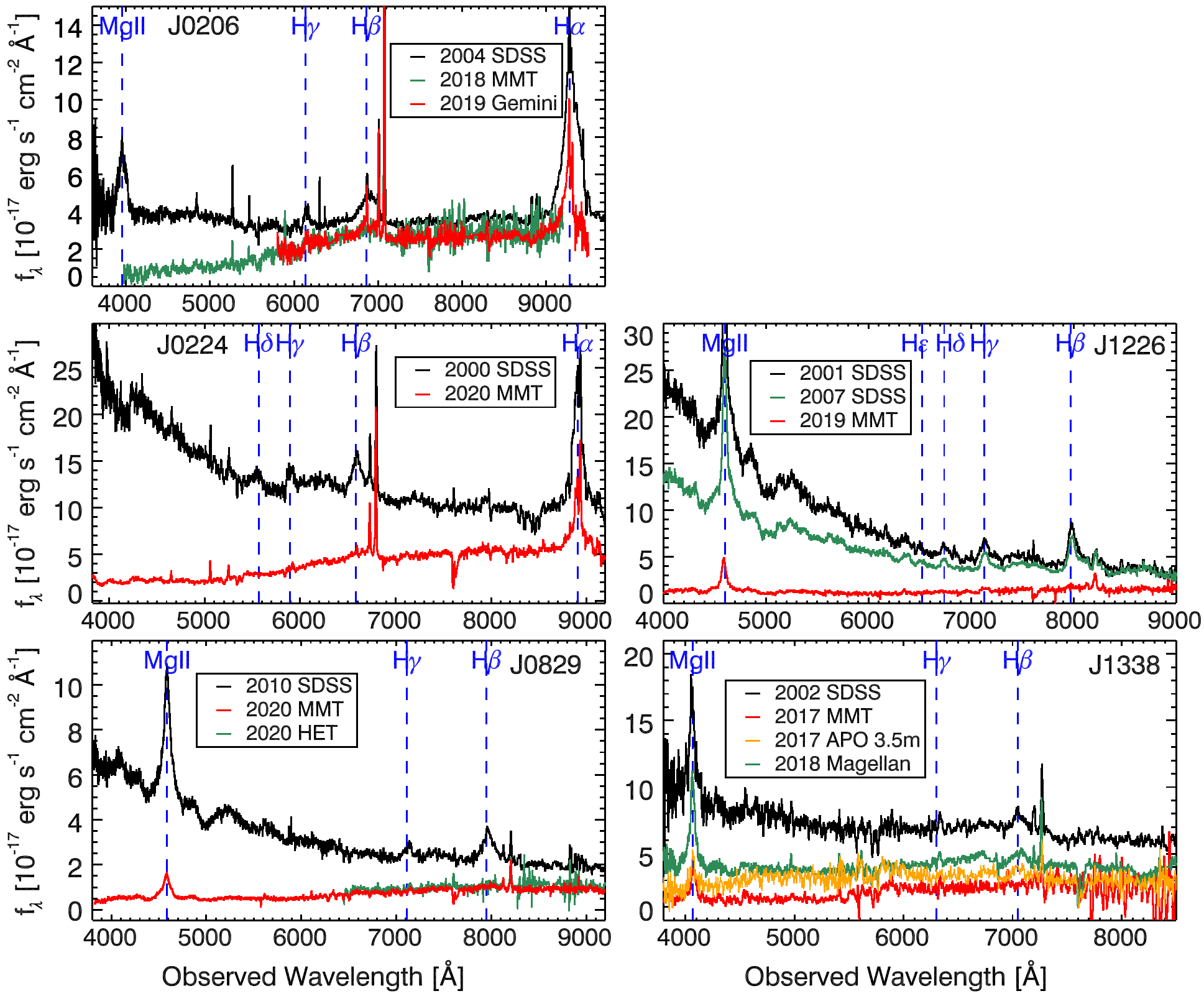}
\caption{Optical spectra for five new CLQs, including J0206 (top left), J0224 (middle left), J0829 (bottom left), J1226 (middle right), and J1338 (bottom right). 
The y-axis is flux density, $f_{\lambda}$, in units of $10^{-17} {\rm erg\,s^{-1}\,cm^{-2}\,\AAm^{-1}}$. The black spectra are from SDSS in bright states. The red spectra are our spectroscopic follow-ups in the faint states. For some CLQs, there are spectra in intermediate states (green and orange). The vertical blue dashed lines mark the expected central wavelengths of broad emission lines. The spectra in the faint states obviously show that the continuum emission faded dramatically, accompanied by the disappearance of broad emission lines.}
\label{fig:optical_spec}
\end{figure*}

\subsection{Optical Spectroscopic Observations} \label{sec:optical_spec}
For the bright state, we use the SDSS spectra, taken by the Sloan Foundation 2.5 m telescope \citep{Gunn2006} at Apache Point Observatory.
The spectra are from the SDSS-I/II, obtained between 2000 to 2014, with a wavelength coverage of 3800-9100\AAm, and the Baryon Oscillation Spectroscopic Survey \citep[BOSS;][]{Dawson2013} spectrograph of the SDSS-III \citep{Eisenstein2011}, with a wider wavelength coverage of 3600-10400\AAm\ \citep{Smee2013}.
The SDSS spectral resolution is 1500 at 3800\AAm\ and 2500 at 9000\AAm.

We obtained optical spectroscopic follow-up for CLQ candidates in the faint states from 2017 to 2020. Table \ref{tab:Optical_obs} summarizes the spectroscopic observation information, including telescope, grating, and exposure time.

The MMT telescope is a 6.5m telescope on Mount Hopkins in Arizona. J1338 was observed with the Blue Channel Spectrograph using a grating 300 lines/mm, with a resolution of 6.47\AAm\ and a wavelength coverage of 5268\AAm. For the other four targets, we used the Binospec Spectrograph with a grating 270 lines/mm with a resolution of 1340 and wavelength coverage of 3900-9240\AAm.

The Gemini Observatory consists of twin 8.1m diameter telescopes, Gemini North and Gemini South, which are located at two separate sites on Maunakea in Hawaii and Cerro Pach\'on in Chile, respectively. We obtained Gemini North observations
for J0206 to cover the H$\beta$ and H$\alpha$ emission lines, simultaneously. We used the Gemini Multi-Object Spectrograph (GMOS-N) with the R400 grating (400 lines/mm), with a resolution of 1918 and wavelength coverage of 4160\AAm.

The 3.5m telescope at the Apache Point Observatory (APO) is located in the Sacramento Mountains in Sunspot in New Mexico.
We used the Double Imaging Spectrometer (DIS), which is a medium dispersion spectrograph with separate red and blue channels. In the blue (red) channel, we used the B400 (R300) grating 400 (300) lines/mm with a wavelength coverage of 3660 (4620) \AAm. Soon after our initial follow-up MMT observations confirmed J1338 to be a CLQ in the faint state, it re-brightened, as confirmed using the APO 3.5m telescope after our {\it Chandra} observations.

The Hobby–Eberly Telescope (HET) is a 10-meter aperture telescope located at the McDonald Observatory in Davis Mountains, Texas. We used the LRS2-R spectrograph, which simultaneously covers from 6450-8400 \AAm\ (Red Arm) and 8275-10500 \AAm\ (Far Red Arm) at a resolving power of 1800 in each channel, respectively \citep{Chonis2016}.

The Magellan Telescopes are a pair of 6.5 meter telescopes located at Las Campanas Observatory in Chile. We used the Inamori Magellan Areal Camera and Spectrograph (IMACS) on the Baade telescope with a grating 300 lines/mm, with a wavelength coverage of 3650-9740\AAm. 
We used the Magellan Baade telescope to observe J1338 a few months after {\it Chandra} observations. 

The spectra were reduced using standard \texttt{IRAF}\footnote{IRAF is distributed by the National Optical Astronomy Observatory, which is operated by the Association of Universities for Research in Astronomy (AURA) under cooperative agreement with the National Science Foundation.} routines \citep{Tody1986, Tody1993}. Since absolute flux calibrations from the follow-up spectra were not always reliable, we normalized the spectra to the earlier SDSS spectrum assuming a constant narrow emission line luminosity (described in \textsection \ref{sec:spec_fit}).

All the spectra of the five CLQs are shown in Figure  \ref{fig:optical_spec}. The spectra in black are the earliest SDSS spectra in the bright states. The spectra in red are our follow-ups in the faint state. For some of them, there are intermediate spectra (green and orange) observed by SDSS or our follow-ups. 

\begin{deluxetable*}{cccccccl}
\tablecaption{Spectroscopic Observations \label{tab:Optical_obs}}
\tablewidth{1pt}
\tablehead{
\colhead{Name} &
\colhead{MJD} &
\colhead{Date} & 
\colhead{Telescope} &
\colhead{Spectrograph} &
\colhead{Grating} & 
\colhead{ExpTime} &
\colhead{State} \\
& day & & & & lines$/$mm & second & 
}
\startdata
J0206 & 56660 & 2014-01-03 & SDSS & $\cdots$ & $\cdots$ & $\cdots$ & Bright \\
J0206 & 58431 & 2018-11-09 & MMT & Binospec & 270 & 2000 & Faint \\
J0206 & 58507 & 2019-01-24 & Gemini & GMOS-N & 400 & 3000 & Faint2 \\
\hline
J0224 & 51908 & 2000-12-30 & SDSS & $\cdots$ & $\cdots$ & $\cdots$ & Bright \\
J0224 & 59145 & 2020-10-23 & MMT & Binospec & 270 & 1800 & Faint \\
\hline
J0829 & 55513 & 2010-11-13 & SDSS & $\cdots$ & $\cdots$ & $\cdots$ & Bright \\
J0829 & 58866 & 2020-01-18 & MMT & Binospec & 270 & 2700 & Faint\\
J0829 & 58968 & 2020-04-29 & HET & LRS2-R & Red 920 (Far Red 800) & 3000 & Faint2 \\
\hline
J1226 & 51990 & 2001-03-22 & SDSS & $\cdots$ & $\cdots$ & $\cdots$ & Bright \\
J1226 & 54153 & 2007-02-22 & SDSS & $\cdots$ & $\cdots$ & $\cdots$  & Intermediate \\
J1226 & 58574 & 2019-04-01 & MMT & Binospec & 270 & 3200 & Faint \\
\hline
J1338 & 52427 & 2002-06-02 & SDSS & $\cdots$ & $\cdots$ & $\cdots$ & Bright \\
J1338 & 57818 & 2017-03-06 & MMT & Blue Channel & 300 & 1800 & Faint \\
J1338 & 57891 & 2017-05-18 & APO 3.5m & DIS & Blue 400 (Red 300) & 900 & Intermediate\\
J1338 & 58162 & 2018-02-13 & Magellan & IMACS & 300 & 1800 & Intermediate2 \\
\enddata
\tablecomments{
Date is in format of year-month-day.
ExpTime stands for ``Exposure Time".
}
\end{deluxetable*}

\subsection{X-ray Observations} \label{sec:Xray_obs}
We used archival X-ray observations close in time to the optical spectra from the {\it Chandra} Source Catalog \citep{Evans2020} or the {\it XMM-Newton} serendipitous survey \citep[4XMM-DR10;][]{Traulsen2020}. 
For X-ray observed CLQ candidates confirmed as CLQs by optical follow-up spectra,  we obtained {\it Chandra} ToO observations. 
Our {\it Chandra} observations were triggered shortly after our optical spectroscopic follow-ups. The targets were observed using the Advanced CCD Imaging Spectrometer (ACIS) detector \citep{Garmire2003}.  The X-ray observations are listed in Table \ref{tab:Xray_obs}.

Four of the CLQ candidates were thereby observed in X-rays in both bright and faint states.  However, J1338 is a special case.
In March 2017 (MJD 57818), we obtained an optical spectrum of the {\it XMM-Newton} source and CLQ candidate J1338, for which analysis showed that the broad H$\beta$ line had vanished at $\geq 3\sigma$ significance with respect to the earlier SDSS spectrum.  Shortly afterward, we began monitoring the source with the 1.2m at FLWO, which confirmed the object to be in the faint state according to the $g$-band magnitude. We triggered the {\it Chandra} observation in mid-April, but it was observed by {\it Chandra} about 3 weeks later, during which we saw an unexpected $\sim$1~mag brightening in the $g$ band. The resulting Chandra spectrum showed a bright unobscured, normal AGN power law slope, as expected after its re-brightening.  Just after the {\it Chandra} observation, we obtained an optical spectrum with the 3.5m at Apache Point Observatory in June that showed significant broad H$\beta$ emission line flux (spectrum plotted in orange in Figure \ref{fig:optical_spec}). 
The light curve of J1338 indicates a transition from faint to bright state of about 40 days, very rapid even for a CLQ.

For the {\it Chandra} data reduction, we used the software \texttt{CIAO} version 4.14 \citep{Fruscione2006} and CALDB version 4.9.6. We used the \texttt{chandra\_repro} reprocessing script, which automates the recommended data processing steps in \texttt{CIAO}.

For the archival {\it XMM-Newton} data reduction, we used the Science Analysis Software \citep[SAS;][]{Gabriel2004} version 20.0.0 following standard procedures.
The {\it XMM-Newton} data is from three X-ray CCD cameras, including two MOS cameras (MOS1 and MOS2) and one pn camera. 

\begin{deluxetable*}{ccccccclcc}
\tablecaption{X-ray Observations \label{tab:Xray_obs}}
\tablewidth{1pt}
\tablehead{
\colhead{Name} &
\colhead{Telescope} &
\colhead{ObsID} &
\colhead{PI} &
\colhead{MJD} &
\colhead{Date} & 
\colhead{Exposure} &
\colhead{Net Counts} & 
\colhead{Group} &
\colhead{Epoch}\\
& & & & & day & kilosecond & & & 
}
\startdata
J0206 & {\it XMM} & 677670138 & Pierre & 55945 & 2012-01-19 & 10.7 & 20.23/18.22/18.22 & 1 & \\
J0206 & {\it XMM} & 677670139 & Pierre & 55945 & 2012-01-19 & 14.3 & 24.63/29.77/18.83 & 1 & \\
J0206 & {\it XMM} & 747190632 & Pierre & 56691 & 2014-02-03 & 24.3 & 170.54/157.70/300.24 & 2 &\\
J0206 & {\it XMM} & 742430201 & Mantz & 56837 & 2014-06-29 & 30.8 & 640.84, 623.45, 436.19 & 3 & 1 \\
J0206 & {\it Chandra} & 16575 & Mantz & 57197 & 2015-06-24 & 4.991 & 36.88  & 4 & \\
J0206 & {\it Chandra} & 20457 & Green & 58483 & 2018-12-31 & 19.84 & 345.86 & 5 & 2 \\
J0206 & {\it Chandra} & 22027 & Green & 58483 & 2018-12-31 & 19.84 & 315.39 & 5 & 2\\
\hline
J0224 & {\it XMM} & 655343850 & Finoguenov & 55582 & 2011-01-21 & 2.7 & 48.72, 58.51, 219.87 & 1 & 1 \\
J0224 & {\it Chandra} & 23662 & Green & 59279 & 2021-03-06 & 24.95 &  357.87 & 2 & 2\\
J0224 & {\it Chandra} & 24978 & Green & 59280 & 2021-03-07 & 23.96 & 389.57 & 2 & 2 \\
\hline
J0829 & {\it XMM} & 724791301 & Dennerl & 56573 & 2013-10-08 & 26.2 & 128.53, 149.15, 175.94 & 1 & 1\\
J0829 & {\it XMM} & 724791332 & Dennerl & 56574 & 2013-10-09 & 8.7 & 89.97, 89.88, 115.24 & 1 & 1 \\
J0829 & {\it XMM} & 724791333 & Dennerl & 56574 & 2013-10-09 & 8.7 & 46.86, 44.57, 46.28 & 1 & 1\\
J0829 & {\it Chandra} & 22550 & Green & 58967 & 2020-04-28 & 17.72 & 77.17 & 2 & 2\\
J0829 & {\it Chandra} & 23234 & Green & 58972 & 2020-05-03 & 29.4 & 124.99 & 2 & 2\\
\hline
J1226 & {\it Chandra} & 4865 & Richards & 53047 & 2004-02-12 & 4.899 & 145.68 & 1 & 1\\
J1226 & {\it Chandra} & 21412 & Green & 58606 & 2019-05-03 & 16.84 & 141.47 & 2 & 2 \\
J1226 & {\it Chandra} & 22204 & Green & 58607 & 2019-05-04 & 17.83 & 134.38 & 2 & 2 \\
J1226 & {\it Chandra} & 22205 & Green & 58608 & 2019-05-05 & 11.91 & 86.96 & 2 & 2 \\
\hline
J1338 & {\it XMM} & 502060101 & Lamastra & 54292 & 2007-07-11 & 17.6 & 68.44, 87.38, 224.18 & 1 & 1 \\
J1338 & {\it Chandra} & 19474 & Green & 57881 & 2017-05-08 & 32.64 & 924.96 & 2 & 2 \\
\enddata
\tablecomments{
ObsID stands for ``Observation ID".
Net counts for {\it XMM-Newton}  ({\it XMM}) are for cameras MOS1, MOS2, and pn, respectively. 
}
\end{deluxetable*}

\begin{figure*}[!ht]
\centering
\includegraphics[width=0.95\textwidth]{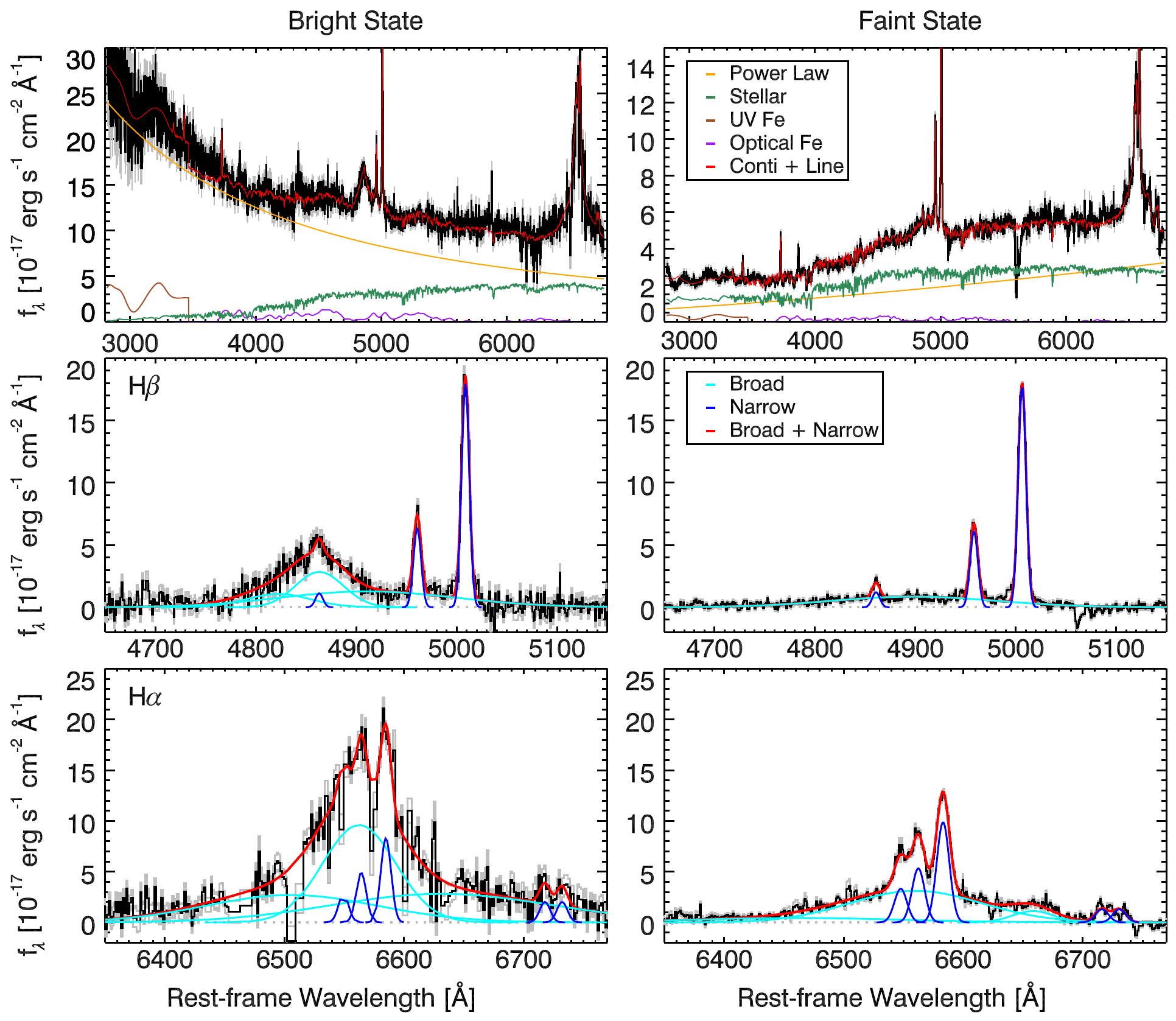}
\caption{Optical spectral fitting to J0224, as an example, in the bright state (left panels) and faint state (right panels). The top panels show the decomposition of continuum emission, including AGN power-law  (orange), host galaxy stellar emission (green), UV \FeII\ emission (brown), and optical \FeII\ emission (purple). The black and gray lines are the spectral flux densities (in units of $10^{-17} {\rm erg\,s^{-1}\,cm^{-2}\,\AAm^{-1}}$) and their uncertainties. The red lines are the total models from the line emission and all the continuum components described above. In the middle (H$\beta$) and bottom (H$\alpha$) panels, there are broad (cyan) and narrow (blue) emission line components.
\label{fig:Spec_fit}}
\end{figure*}

\subsection{Radio Observations} \label{sec:radio_obs}

If CLQ dim states represent a radiatively inefficient accretion (RIAF) state, they should re-activate a compact radio jet, as seen in low-luminosity AGN \citep{Ho2008}. We triggered radio observations using the VLA for four of the five CLQs, omitting J1338, since it had returned rapidly to the bright state.  The VLA consists of 27 independent antennas, each of which has a dish diameter of 25 meters \citep{Thompson1980, Perley2011}.  The VLA observations were performed within one day of the {\it Chandra} X-ray observations. 

Each VLA observation lasted two hours, except for J1226 which lasted one hour.  This yielded 87 min on source (38 min for J1226). Observations were taken in C-band with 2$\times$2 GHz basebands centered at 5.25 and 7.2 GHz for the near-equatorial sources J0206, J0224, and J1226 (to minimize interference from satellites within the Clarke Belt), and centered at 5.0 and 7.0 GHz for J0829.  We set the flux density scale and performed bandpass calibrations using 3C 147 (for J0206 and J0224) or 3C 286 (for J0829 and J1226), and we cycled to a nearby phase calibrator to solve for the complex gain solutions (see Table~\ref{tab:Radio_obs} for the phase calibrator and array configuration used for each observation).  

The VLA data were processed using the Common Astronomy Software Applications v6.1.2.7 \citep[CASA;][]{CASA2022} and calibrated using the VLA calibration pipeline 6.1.2, following standard procedures. The data were imaged using {\tt tclean}, adopting two Taylor terms to model the spectral dependence of sources in the field, and adopting Briggs weighting with a {\tt robust} value for each source as noted in Table \ref{tab:Radio_obs}.  We detected radio emission at the known optical position of each target, and flux densities were  measured using {\tt imfit} by fitting a two-dimensional Gaussian to each source.  
We found that J0224 and J1226 are consistent with point sources, and J0206 and J0829 are  extended (\SI{3.0}{\arcsecond}$\times$\SI{0.5}{\arcsecond} and \SI{6.8}{\arcsecond}$\times$\SI{4.5}{\arcsecond}, respectively, after deconvolution with the synthesized beam). 
The signal-to-noise ratio of the radio detections is not large enough to obtain meaningful in-band constraints on the radio spectral indices of our targets. We summarize the radio observations in Table \ref{tab:Radio_obs}.

\begin{deluxetable*}{cccccccccc}
\tablecaption{Radio Observations \label{tab:Radio_obs}}
\tablewidth{1pt}
\tablehead{
\colhead{Name} &
\colhead{Project} &
\colhead{Config.} &
\colhead{Phase Cal.} &
\colhead{MJD} &
\colhead{Date} &
\colhead{Robust} &
\colhead{$f_{peak}$} &
\colhead{log$L_{\rm 5GHz}$} &
\colhead{log($R$)}\\
& & & & & & & $\mu$Jy beam$^{-1}$ & erg s$^{-1}$
}
\startdata
J0206 & 18B-393 & C & J0239$-$0234 & 58483 & 2018-12-31 & 0.0 & 47.1$\pm$5.4 & 39.01 & -4.79\\
J0224 & SM0081 & A & J0239$-$0234 & 59279 & 2021-03-06 & 0.0 & 50.1$\pm$4.5 & 38.90 & -5.03\\
J0829 & SL0102 & C & J0818+4222   & 58968 & 2020-04-29 & 1.0 & 13.8$\pm$2.8 & 38.86 & -5.25\\
J1226 & SK0063 & B & J1224+0330   & 58609 & 2019-05-06 & 0.5 & 28.4$\pm$5.2 & 39.18 & -5.15\\
\enddata
\tablecomments{
$f_{peak}$ is reported at 6.2 GHz for J0206, J0224, and J1226, and 6.0 GHz for J0829.  $R$ is the radio loudness, $R = L_{\rm 5GHz}/L_{2500\AAm}$.
}
\end{deluxetable*}

\subsection{Infrared Data}
The Wide-field Infrared Survey Explorer \citep[WISE;][]{Wright2010} mapped the all-sky in 2010 in four bands centered at mid-infrared (MIR) wavelengths of 3.4, 4.6, 12, and 22 $\mu$m ($W1$, $W2$, $W3$, and $W4$). WISE scans the sky every half-year in the $W1$ and $W2$ bands. We used unWISE \citep{Lang2014,Meisner2023} light curves force-photometered for sources in the DESI Legacy Imaging Surveys \citep{Dey2019}. 

We also compiled available near-infrared (NIR) data, including the UKIRT Infrared Deep Sky Survey \citep[UKIDSS;][]{Lawrence2007} in the $YJHK$ bands, the UKIRT Hemisphere Survey \citep[UHS;][]{Dye2018} in the $J$ band, and the VISTA Hemisphere Survey \citep[VHS;][]{McMahon2013} in the $YJHK_s$ bands.

\begin{figure*}[!ht]
\centering
  \subfigure{
  \includegraphics[width=0.45\textwidth]{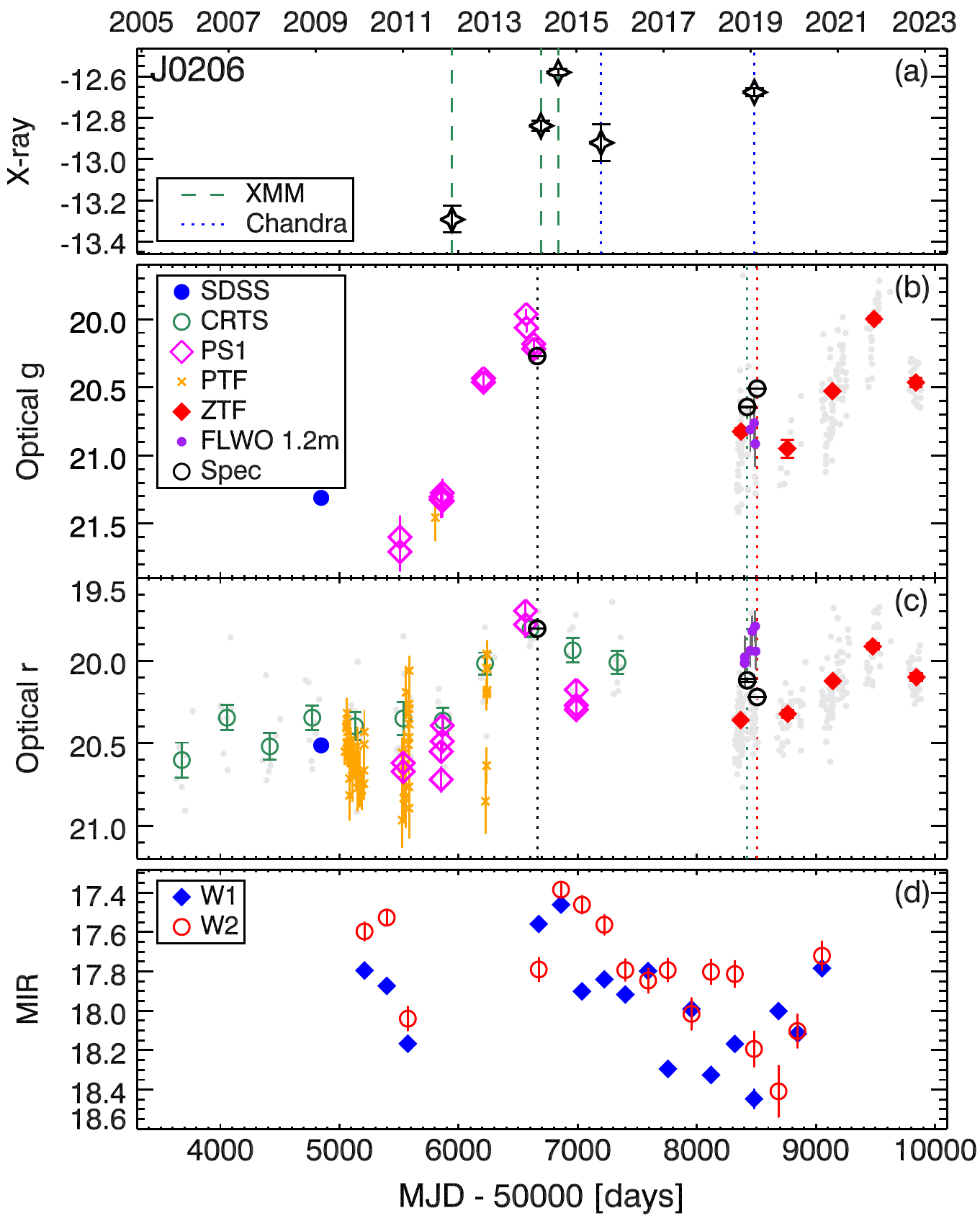}}
\subfigure{
  \includegraphics[width=0.45\textwidth]{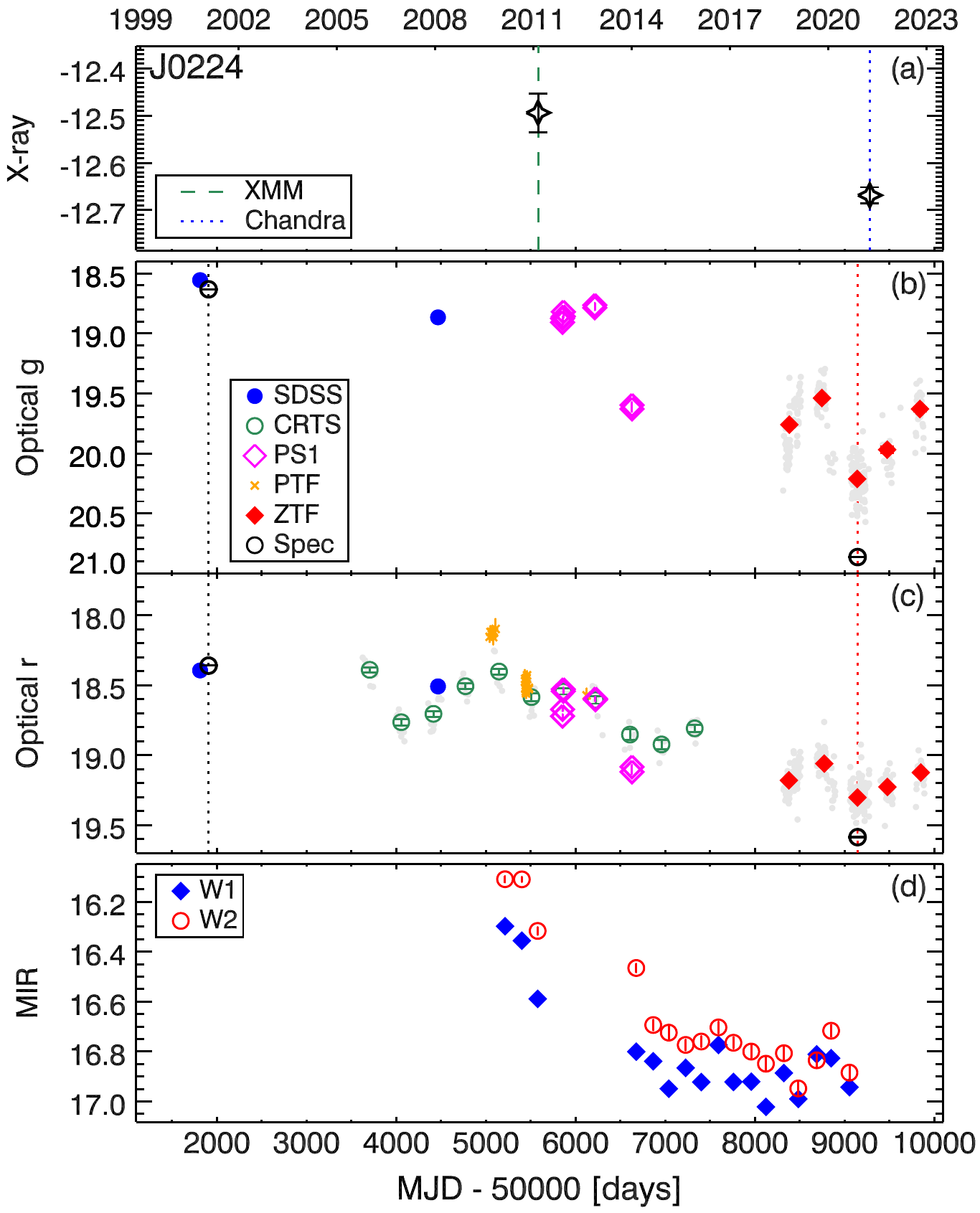}}\\
\vspace{-0.2cm}
  \subfigure{
  \includegraphics[width=0.45\textwidth]{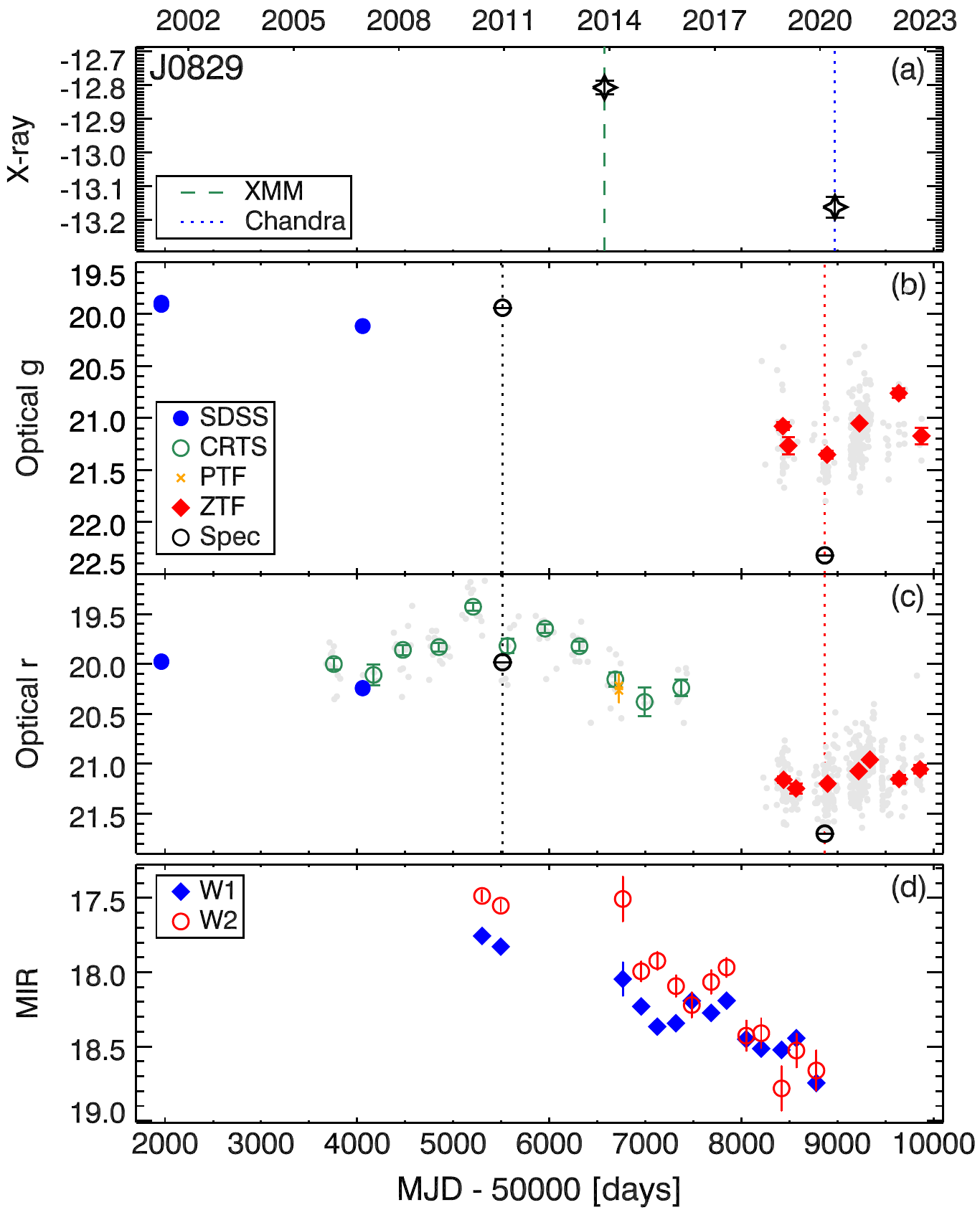}}
\subfigure{
  \includegraphics[width=0.45\textwidth]{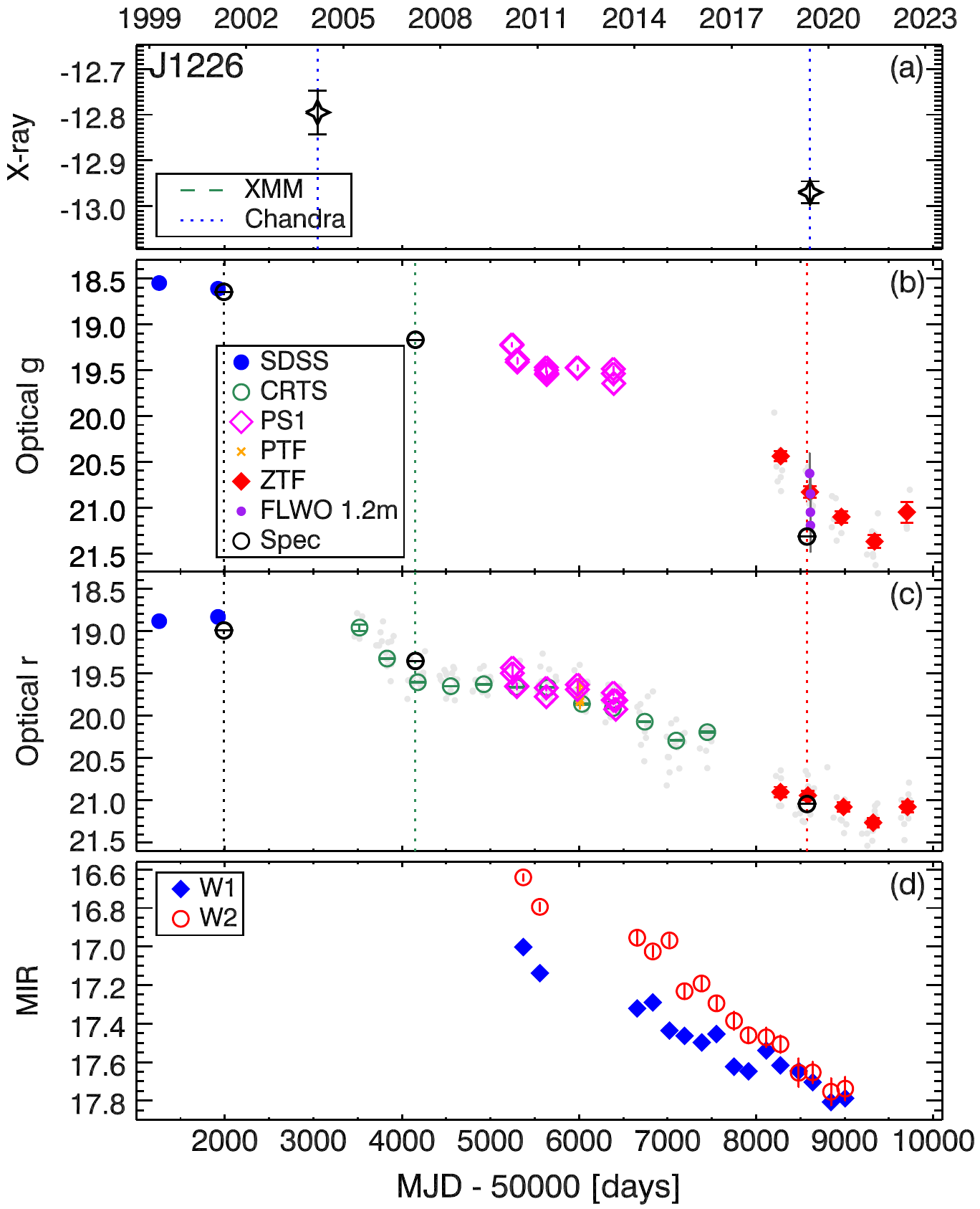}}\\
  \caption{Light curves of the CLQs. The X axis is MJD - 50000 days, and the X axis across the top indicates the first date in each year.
For each CLQ, the five panels are (a) X-ray 0.5 - 7 keV flux (${\rm erg~s^{-1}~cm^{-2}}$) in logarithm, (b) optical $g$-band magnitude, (c) optical $r$-band magnitude, and (d) WISE MIR $W1$-band (blue) and $W2$-band (red) Vega magnitude. 
In panel (a), the vertical lines indicate the epochs of X-ray observations, including {\it XMM} (green; dashed) and {\it Chandra} (blue; dotted). 
In panel (b) and (c), the different colors and symbols are from various optical imaging surveys, including SDSS (blue filled circles), CRTS (gray dots, with the mean for each year's data as green open circles), PS1 (magenta open diamonds), PTF (orange crosses), ZTF (gray dots, with the mean for each year's data as red filled diamonds), FLWO (purple dots), and synthetic photometry from spectra (black open circles).
All optical magnitudes have been calibrated to SDSS magnitudes.
The vertical dotted lines indicate the epochs of optical spectroscopic observations. 
The multiwavelength observations obviously show the same variability trend. Recent ZTF and/or FLWO data shows signs of re-brightening. \label{fig:LC}}
\end{figure*}

\begin{figure}
\centering
\includegraphics[width=0.45\textwidth]{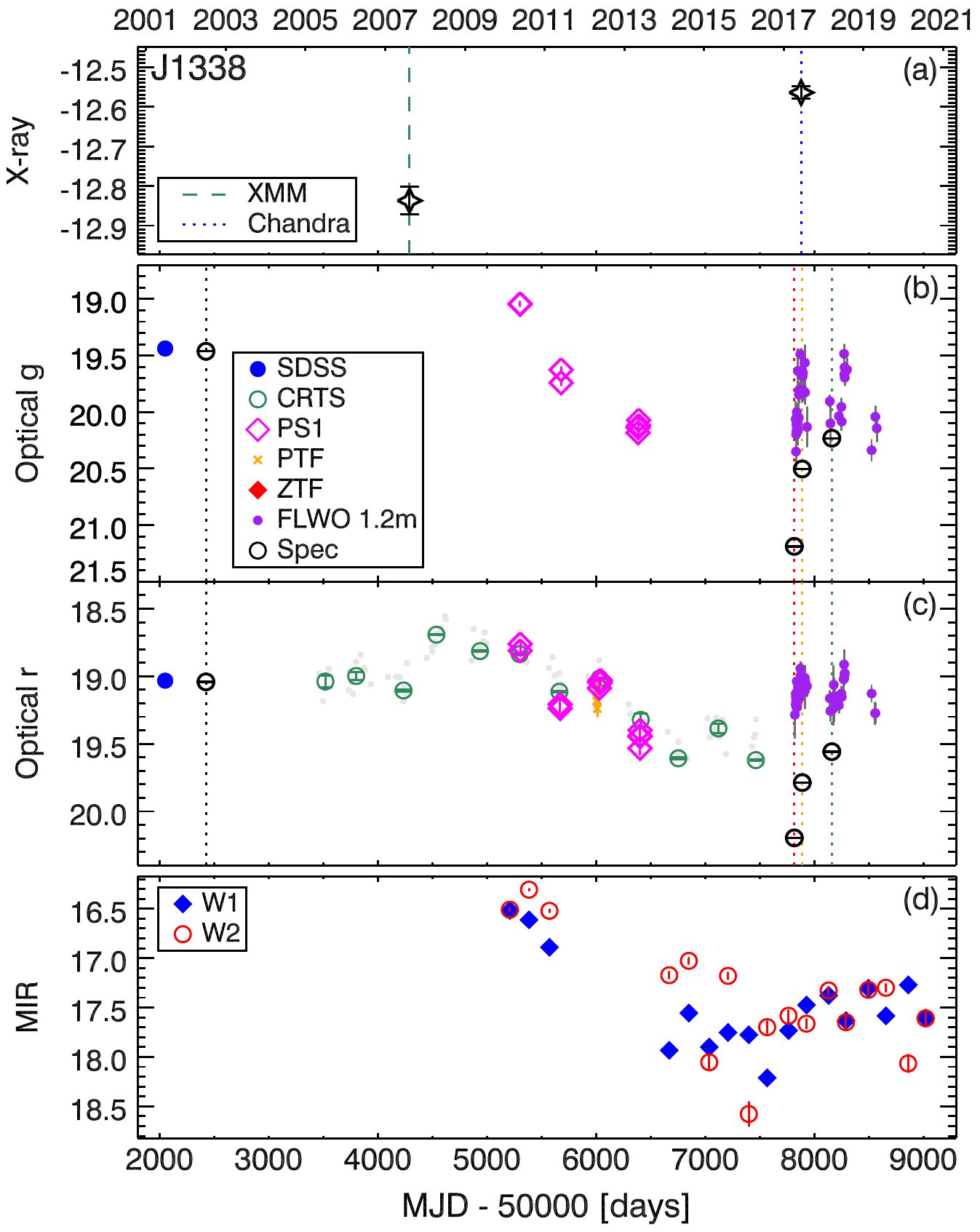}
\caption{Same as Figure \ref{fig:LC}.
\label{fig:LC_one}}
\end{figure}

\section{Data Analysis}

\subsection{Optical Spectral Fitting} \label{sec:spec_fit}
We fit the optical spectra following the quasar spectral fitting code, \texttt{QSOfit} \citep{Shen2019}. The spectra are fit in the rest frame of the quasar after correcting for Galactic reddening, again using the dust map of \citet{Schlegel1998} and the extinction curve from \citet{Cardelli1989}. 

\texttt{QSOfit} decomposes different components in the quasar spectrum, including power-law continuum, \FeII\ emission multiplets, and major broad and narrow emission lines. It uses empirical UV \FeII\ emission templates from the literature \citep{Vestergaard2001, Tsuzuki2006, Salviander2007} covering from 1000 to 3500\AAm, and an optical \FeII\ template (3686-7484\AAm) from \citet{Boroson1992}. 

\texttt{QSOfit} is designed for quasar spectral fitting, where the host galaxy emission is not prominent, so no host galaxy components were included initially. For our CLQs in the faint states when the AGN continuum is dim, the host galaxy emission becomes more obvious. Therefore, we enhanced the code to take into account host galaxy stellar emission using the simple stellar population models \citep{Bruzual2003} with the \citep{Chabrier2003} initial mass function. We allow for two host components, one for the young ($<$300 Myr) stellar population and one for old ($>$300 Myr) stellar populations. We use 30 templates from \citet{Bruzual2003}, covering metallicities of $Z$ = 0.004, 0.02 and 0.05, and the ages of young (0.005, 0.025, 0.10, and 0.29 Gyr) and old (0.64, 0.90, 1.4, 2.5, 5, and 11 Gyr) populations. 

Following \citet{Shen2019}, we choose a few continuum windows and fit the continuum components described above together. The emission lines are fit after subtracting the fitted continuum model. Three broad and one narrow Gaussian components are used for the H$\beta$ and H$\alpha$ emission. 

According to the empirical $L_{\mathrm [OIII]} - R_{\mathrm NLR}$ relation of \citet{Bennert2002}, 
the size of the \OIII\ narrow emission line region for our CLQs are at most 
$2-3$\,kpc corresponding to $\sim 10^4$ light-years, so the narrow-line flux is certainly expected to remain constant on decadal timescales. The SDSS fiber sizes are \SI{2}-\SI{3}{\arcsecond}, and our follow-up spectral slit widths are \SI{1}{\arcsecond}.
At the redshifts of our CLQ sample, the scale is 7.7 kpc/\SI{}{\arcsecond}, so the NLRs are fully encompassed by the spectral apertures, even accounting for seeing. 
We use the first SDSS epoch as the reference spectrum and rescale the other spectra to the first SDSS epoch according to the flux of the narrow \OIII, and then fit the spectra as described above.

To quantify the measurement uncertainties, we used a Monte Carlo approach, by adding a random Gaussian deviate to the flux at each pixel, with the Gaussian $\sigma$ equal to the spectral error at that pixel. We compute the values and their uncertainties using the median value and the semi-amplitude of the range enclosing the 16th and 84th percentiles of the distribution from 100 trials. 

Figure \ref{fig:Spec_fit} shows an example of the optical spectral fitting to J0224 in the bright state (left panels) and faint state (right panels). The top panels show the continuum decomposition. In the bright state, the continuum is dominated by the AGN power-law continuum. Weak stellar emission from the host is also detected. In the faint state, the AGN continuum emission dimmed dramatically, thus the continuum becomes dominated by the host stellar emission. 
The middle two panels show the spectral fitting in the H$\beta$ window. There is obvious broad H$\beta$ emission in the bright state; but it becomes invisible in the faint state. The bottom two panels are in the H$\alpha$ window. The broad H$\alpha$ emission becomes much weaker, but remains visible in the faint state.

To evaluate the variability of broad H$\beta$ emission, we calculate the flux deviation between the bright- and faint-state spectra in the H$\beta$ window (4750–4940\AAm) following \citet{MacLeod2019} and \citet{Green2022} as
\begin{equation}\label{eq:Nsig}
N_{\sigma}(\lambda) = (f_{\rm bright} - f_{\rm faint}) / \sqrt{\sigma_{\rm bright}^2 + \sigma_{\rm faint}^2},
\end{equation}
where $f$ is the flux density in units of erg s$^{-1}$ cm$^{-2}$ $\AAm^{-1}$ and $\sigma$ is its uncertainty. Following the procedures in \citet{MacLeod2019} and \citet{Green2022}, we subtract the continuum model, re-bin the spectrum to $2\AAm/$pixel in the rest frame, and smooth the spectrum with a window of $32\AAm$\footnote{We also tested a smaller spectral smoothing window of $16\AAm$, and the calculated $N_{\sigma}({\rm H}\beta)$ values are slightly larger ($\sim10\%$) than those using $32\AAm$.}. 

We characterize the significance of the H$\beta$ change as the maximum flux deviation of the line relative to the continuum at $4750\AAm$ as follows:
\begin{equation}\label{eq:Nsig_Hb}
N_{\sigma}({\rm H}\beta) = N_{\sigma} (4750-4940 \AAm) - N_{\sigma} (4750 \AAm).
\end{equation}

To estimate the black hole mass, $M_{\rm BH}$, for each of our CLQs, we utilize the single-epoch virial black hole mass method, using the bright-state spectrum with its prominent broad emission lines. In this approach, the $M_{\rm BH}$ is estimated using the measured width of a broad Balmer emission line, as well as an empirical radius–luminosity ($R-L$) relation for the BLR based on reverberation mapping results of low-redshift AGN. We use the relation (Equation 6) in \citet{Vestergaard2006} as follows,
\begin{multline}\label{eq:MBH}
M_{\rm BH} = 10^{6.91 \pm 0.02} \left[ \frac{{\rm FWHM(H}\beta)}{\rm 1000~km~s^{-1}} \right]^2 \\ 
\times \left[ \frac{L({\rm H}\beta)}{\rm 10^{42}~erg~s^{-1}} \right]^{0.5} M_{\odot},
\end{multline}
where ${\rm FWHM(H}\beta)$ and $L({\rm H}\beta)$ are the full width at half maximum (FWHM; i.e., the width) and luminosity of the broad H$\beta$ emission. We calculate the Eddington luminosity using $L_{\rm Edd} = 1.26 \times 10^{38} M_{\rm BH}$, in units of erg s$^{-1}$.

\subsection{X-ray Spectral Analysis} \label{sec:xray_fit}

For X-ray spectral analysis and fitting, we used the \texttt{XSPEC} software \citep{Arnaud1996} in version 12.12.0. We grouped X-ray spectra observed with the same telescope within one week and performed spectral analysis together, indicated by the column ``Group" in Table \ref{tab:Xray_obs}. In \texttt{XSPEC}, the spectra are re-binned using the task \texttt{grppha}, so that each spectral bin contains 20, 15, 10, 5, 3 or 1 counts, when the total net counts, ${N_{\rm net}}$, ${N_{\rm net}}\geq300$, $225\leq {N_{\rm net}} <300$, $100\leq {N_{\rm net}} <225$, $50\leq {N_{\rm net}} <100$, $30\leq {N_{\rm net}} <50$, or $10\leq {N_{\rm net}} <30$, respectively. We used $\chi^2$ statistic for high-count spectra and C statistic \citep{Cash1979} when the counts in each spectral bin are less than 20.

The intrinsic AGN X-ray spectrum is typically a power law within the {\it Chandra} and {\it XMM-Newton} energy range. The observed X-ray spectrum is a result of the intrinsic spectrum modified by line-of-sight absorption, mainly from the intrinsic absorption at the quasar redshift and the Galactic absorption from the Milky Way. Therefore, we used the the following \texttt{XSPEC} model,
\begin{equation}\label{eq:model2}
{\rm phabs} * {\rm zphabs} * {\rm zpower},
\end{equation}
where \texttt{phabs} is the Galactic absorption, \texttt{zphabs} models
the intrinsic absorption with the redshift fixed to the source redshift, and \texttt{zpower} is a simple photon power law redshifted to the source redshift. We used the \texttt{nh} task in \texttt{HEAsoft} to obtain the Galactic absorption column density, ${\rm N_H}$, derived  from  the  HI map \citep{Kalberla2005}. The typical Galactic values ${\rm N_H}$ for our CLQs are $2\times10^{20}~{\rm cm}^{-2}$. 

\subsection{SED Fitting} \label{sec:SED_fit}
We constructed SEDs for each object separately for the bright and faint states by using the photometric and spectroscopic measurements closest to those states.
Since we are interested in the Eddington ratio in different states for these accreting SMBHs, we calculate the bolometric luminosity, $L_{\rm bol}$, by fitting to the SEDs. We use the \texttt{X-CIGALE} software \citep{YangGuang2020} version v2022.1. \texttt{X-CIGALE} implements a modern UV-to-IR AGN module, SKIRTOR \citep{Stalevski2012, Stalevski2016}, which is a clumpy torus model. Host galaxy stellar emission is also considered in \texttt{X-CIGALE} using stellar population spectra of \citet{Bruzual2003}.

To fit X-ray data into the full SED fit across all wavelengths, \texttt{X-CIGALE} uses the UV to X-ray slope, i.e., $\alpha_{\rm OX}$. We use a definition of $\alpha_{\rm OX}$ following the literature \citep[e.g.,][]{Tananbaum1979, Lusso2010, Ruan2019, Jin2021} as, 
\begin{equation}
    \alpha_{\rm OX} = 0.3838~{\rm log}(L_{\rm 2500\AAm} / L_{\rm 2~keV}),
\end{equation}
where $L_{\rm 2500\AAm}$ and $L_{\rm 2~keV}$ are the AGN intrinsic luminosities per frequency (in units of erg s$^{-1}$ Hz $^{-1}$) at 2500\AAm\ (UV) and 2 keV (X-ray), respectively. 

Since we have X-ray measurements as described in \textsection \ref{sec:xray_fit}, we fixed the X-ray photon indices, $\Gamma_{\rm X}$ and normalizations (fluxes), to the values measured from the X-ray spectra. To fit the full SED across undetected regions such as the EUV, \texttt{X-CIGALE} uses expected AGN values of $\alpha_{\rm OX} = 0.137~{\rm log}(L_{\rm 2500\AAm}) - 2.638$ \citep{Just2007}, within a specified tolerance. To explore a wide range of SED shapes and possible variability of $\alpha_{\rm OX}$, we allow for a large deviation $|\Delta \alpha_{\rm OX}|_{\rm max}=0.6$ from this relation. 
The calculated $\alpha_{\rm OX}$ values are fairly robust to the fact that the photometric data and X-ray observations were not taken simultaneously.  For example, a change of 30\% in $L_{\rm 2500\AAm}$ or $L_{\rm 2~keV}$ leads to a difference of 0.04 in $\alpha_{\rm OX}$, which is comparable to the uncertainties for $\alpha_{\rm OX}$.

Galactic extinction correction is not implemented in \texttt{X-CIGALE}. For optical, NIR, and MIR data, we corrected the Galactic extinction for each band as described in \textsection \ref{sec:optical_imaging}. 

\movetabledown=2.5in
\begin{rotatetable*}
\begin{deluxetable*}{ccccccccccccccc}
\setlength{\tabcolsep}{1pt}
\tablecaption{Optical Spectroscopic Properties \label{tab:Optical_prop}}
\tablewidth{1pt}
\tablehead{
\colhead{Name} &
\colhead{$N_\sigma({\rm H}\beta)$} &
\colhead{log($L_{{\rm H} \beta, 1}$)} &
\colhead{log($L_{{\rm H} \beta, 2}$)} &
\colhead{log($L_{{\rm H} \alpha, 1}$)} &
\colhead{log($L_{{\rm H} \alpha, 2}$)} &
\colhead{log($L_{\rm MgII, 1}$)} &
\colhead{log($L_{\rm MgII, 2}$)} &
\colhead{log($L_{\rm 3000, 1}$)} &
\colhead{log($L_{\rm 3000, 2}$)} &
\colhead{log($\lambda L_{ 5100 \AAm}$)} &
\colhead{log($L_{\rm [OIII]}$)} &
\colhead{FWHM$_{\rm H\beta}$} &
\colhead{log$(M_{\rm BH})$} &
\colhead{log$(M_{\rm BH, ref})$} \\
& & erg s$^{-1}$ & erg s$^{-1}$ & erg s$^{-1}$ & erg s$^{-1}$ & erg s$^{-1}$ & erg s$^{-1}$ & erg s$^{-1}$ & erg s$^{-1}$ & erg s$^{-1}$ & erg s$^{-1}$ & km s$^{-1}$ & \(M_\odot\) & \(M_\odot\)
}
\startdata
J0206 & 3.59 & $42.32\pm0.02$ & $41.84\pm0.12$ & 43.04$\pm$0.01 & 42.66$\pm$0.01 & 42.24$\pm$0.02 & $\cdots$ & $44.03\pm0.01$ & $43.70\pm0.03$ & $43.06\pm0.01$ & 42.14 &$9700\pm500$ & $8.90\pm0.05$ & $8.95\pm0.04$ \\
J0224 & 5.85 & $42.50\pm0.01$ & $42.01\pm0.02$ & 43.04$\pm$0.03 & 42.53$\pm$0.02 & $\cdots$ & $\cdots$ & $44.38\pm0.01$ & $44.58\pm0.01$ & $43.14\pm0.09$ & 42.00 & $5100\pm300$ & $8.51\pm0.05$ & $8.74\pm0.05$ \\
J0829 & 3.91 & $42.67\pm0.05$ & $42.20\pm0.04$ & $\cdots$ & $\cdots$ & 43.02$\pm$0.01 & 42.14$\pm$0.04 & $44.32\pm0.02$ & $44.52\pm0.02$ & $42.90\pm0.02$ & 41.55 & $3900\pm800$ & $8.25\pm0.17$ & $8.54\pm0.13$ \\
J1226 & 4.41 & $43.07\pm0.07$ & $42.38\pm0.03$ & $\cdots$ & $\cdots$ & 43.48$\pm$0.02 & 42.79$\pm$0.01 & $44.60\pm0.02$ & $45.04\pm0.02$ & $43.67\pm0.02$ & 41.97 & $2900\pm600$ & $8.13\pm0.15$ & $8.17\pm0.12$ \\
J1338 & 2.76 & $42.51\pm0.08$ & $42.17\pm0.05$ & $\cdots$ & $\cdots$ & 42.67$\pm$0.05 & 44.24$\pm$0.09 & 43.28$\pm$1.06 & 41.57$\pm$0.09 & $44.30\pm0.05$ & 41.85 & $4400\pm700$ & $8.34\pm0.13$ & $8.60\pm0.15$ \\
\enddata
\tablecomments{
\hspace*{\fill}\begin{minipage}{1.27\textwidth}
$N_\sigma({\rm H}\beta)$ is the significance of the change in H$\beta$ broad-line emission between the bright and faint states. $L_{{\rm H} \beta, 1}$ and $L_{{\rm H} \beta, 2}$ are the broad H$\beta$ line luminosities in the bright and faint state, respectively. $L_{ 5100 \AAm}$ is listed for the bright state only, as that is what we use to calculate $M_{\rm BH}$. $L_{\rm [OIII]}$ is the best-fit rescaled narrow \OIIIb\ line luminosity. $M_{\rm BH, ref}$ are the values derived by \citet{Wu2022}, shown for comprison.
\end{minipage}
}
\end{deluxetable*}
\end{rotatetable*}

\begingroup
\renewcommand*{\arraystretch}{1.2}

\begin{deluxetable*}{cccrrrrrrrrr}
\tablecaption{X-ray Spectral Analysis \label{tab:Xray_model}}
\tablewidth{1pt}
\tablehead{
\colhead{Name} &
\colhead{Telescope} &
\colhead{Epoch} &
\colhead{$\chi^2_{\rm}$} &
\colhead{dof$_{\rm}$} &
\colhead{$\chi^2_{\rm red}$} &
\colhead{${\rm N_{H, intrin}}$} &
\colhead{$\Gamma_{X}$} &
\colhead{Norm} & 
\colhead{${\rm log(F}_{\rm 0.5-7keV})$} &  
\colhead{${\rm log(\nu L_{\rm 2 keV}})$} & 
\colhead{${\rm log(L_{\rm 2-10 keV}})$} \\ 
& & & & & & $10^{22}~{\rm cm}^{-2}$ & & $10^{-5}$ & ${\rm erg~s^{-1}~cm^{-2}}$ & ${\rm erg~s^{-1}}$ & ${\rm erg~s^{-1}}$
}
\startdata
J0206 & {\it XMM-Newton} & 1 & 132.39 & 85 & 1.56 & 0.06$_{-0.01}^{+0.01}$ & 1.74$_{-0.06}^{+0.06}$ & 10.13$_{-0.58}^{+0.63}$ & $-12.58_{-0.02}^{+0.02}$ & $43.79_{-0.01}^{+0.01}$ & $44.08_{-0.02}^{+0.02}$ \\
J0206 & {\it Chandra} & 2 & 22.65 & 29 & 0.78 & 0.00$_{-0.00}^{+0.07}$ & 1.67$_{-0.08}^{+0.09}$ & 7.20$_{-0.63}^{+0.79}$ & $-12.68_{-0.02}^{+0.02}$ & $43.66_{-0.02}^{+0.02}$ & $43.98_{-0.03}^{+0.03}$ \\
\hline
J0224 & {\it XMM-Newton} & 1 & 80.20 & 45 & 1.78 & 0.04$_{-0.02}^{+0.03}$ & 2.34$_{-0.16}^{+0.17}$ & 20.61$_{-2.22}^{+2.65}$ & $-12.49_{-0.04}^{+0.04}$ & $43.81_{-0.03}^{+0.03}$ & $43.89_{-0.06}^{+0.06}$ \\
J0224 & {\it Chandra} & 2 & 54.73 & 35 & 1.56 & 0.00$_{-0.00}^{+0.03}$ & 1.71$_{-0.09}^{+0.09}$ & 7.20$_{-0.68}^{+0.73}$ & $-12.67_{-0.02}^{+0.02}$ & $43.53_{-0.02}^{+0.02}$ & $43.84_{-0.03}^{+0.03}$ \\
\hline
J0829 & {\it XMM-Newton} & 1 & 273.54 & 170 & 1.61 & 0.09$_{-0.03}^{+0.03}$ & 1.90$_{-0.08}^{+0.09}$ & 9.67$_{-0.88}^{+0.98}$ & $-12.81_{-0.02}^{+0.02}$ & $44.05_{-0.02}^{+0.02}$ & $44.29_{-0.03}^{+0.02}$ \\
J0829 & {\it Chandra} & 2 & 45.52 & 27 & 1.69 & 0.00$_{-0.00}^{+0.37}$ & 1.72$_{-0.12}^{+0.22}$ & 3.23$_{-0.55}^{+1.30}$ & $-13.16_{-0.03}^{+0.03}$ & $43.63_{-0.04}^{+0.04}$ & $43.93_{-0.04}^{+0.04}$ \\
\hline
J1226 & {\it Chandra} & 1 & 5.10 & 8 & 0.64 & 0.00$_{-0.00}^{+0.22}$ & 2.19$_{-0.19}^{+0.19}$ & 13.02$_{-2.06}^{+2.37}$ & $-12.80_{-0.05}^{+0.05}$ & $44.10_{-0.04}^{+0.04}$ & $44.23_{-0.07}^{+0.07}$ \\
J1226 & {\it Chandra} & 2 & 61.52 & 46 & 1.34 & 0.00$_{-0.00}^{+0.03}$ & 1.66$_{-0.11}^{+0.11}$ & 4.63$_{-0.58}^{+0.65}$ & $-12.97_{-0.02}^{+0.02}$ & $43.80_{-0.03}^{+0.03}$ & $44.13_{-0.03}^{+0.03}$ \\
\hline
J1338 & {\it XMM-Newton} & 1 & 79.54 & 65 & 1.22 & 0.11$_{-0.03}^{+0.04}$ & 2.25$_{-0.15}^{+0.16}$ & 10.38$_{-1.29}^{+1.56}$ & $-12.84_{-0.04}^{+0.03}$ & $43.72_{-0.03}^{+0.03}$ & $43.83_{-0.05}^{+0.05}$ \\
J1338 & {\it Chandra} & 2 & 42.38 & 35 & 1.21 & 0.00$_{-0.00}^{+0.04}$ & 2.04$_{-0.07}^{+0.07}$ & 14.55$_{-1.01}^{+1.05}$ & $-12.56_{-0.02}^{+0.02}$ & $43.93_{-0.02}^{+0.02}$ & $44.11_{-0.03}^{+0.03}$ \\
\enddata
\tablecomments{
${\rm N_{H, intrin}}$ is the best-fit intrinsic absorption column density at the quasar redshift in units of $10^{22}~{\rm cm}^{-2}$.}
\end{deluxetable*}
\endgroup

\section{Results} \label{sec:results}
\subsection{Optical Results}

The optical spectroscopic observations demonstrate that the quasar continuum emission dimmed significantly, along with the broad Balmer emission. In all five CLQs, broad H$\beta$ and H$\gamma$ disappeared. In three CLQs, J0224, J1226, and J1338, there is visible broad H$\delta$ emission in the bright state, which disappeared in the faint state as well. For two CLQs at lower redshift, thus with coverage of H$\alpha$ emission, J0206 and J0224, the broad H$\alpha$ becomes weaker in the faint state, but still remains visible (see the spectral fitting to J0224 in Figure \ref{fig:Spec_fit}). For the other three CLQs at higher redshift, thus with coverage of \MgII, the broad \MgII\ emission decreases, but remains visible as well. 
This phenomenon is consistent with previous spectroscopic follow-up of CLQs \citep[e.g.,][]{Yang2018, MacLeod2019, Green2022}. 

We calculate the significance of H$\beta$ change, defined in Equation \ref{eq:Nsig_Hb}. Four of the five CLQs vary at $\gtrsim 4\sigma$ level; and one CLQ J1338, the first one observed in 2017 which rapidly brightened again, shows a decrease of broad H$\beta$ slightly below the $3\sigma$ level.

\citet{MacLeod2019} and \citet{Green2022} used a criterion defining CLQ behavior as having a change of broad H$\beta$ flux at $N_{\sigma}({\rm H}\beta) \geq 3$. \citet{Green2022} suggested that additional intrinsic criteria less dependent on S/N might be that the fractional change in both continuum luminosity and broad H$\beta$ line luminosity should be larger than 30\%.  In practice, nearly all CLQs that satisfy the $N_{\sigma}({\rm H}\beta) \geq 3$ criterion also satisfy the other two criteria. In Table \ref{tab:Optical_prop}, we summarize some of the relevant spectral fitting measurements. 
We measure the broad-emission line luminosities (for H$\beta$, H$\alpha$, and \MgII\,) and the rest-frame 3000\AAm\ luminosities, $L_{\rm 3000 \AAm}$, in the bright and faint states. The broad-emission line luminosities and $L_{\rm 3000 \AAm}$ change by more than 30\% between the bright and faint states for all the five CLQs.

\begin{figure*}[!ht]
\centering
\includegraphics[width=1.0\textwidth]{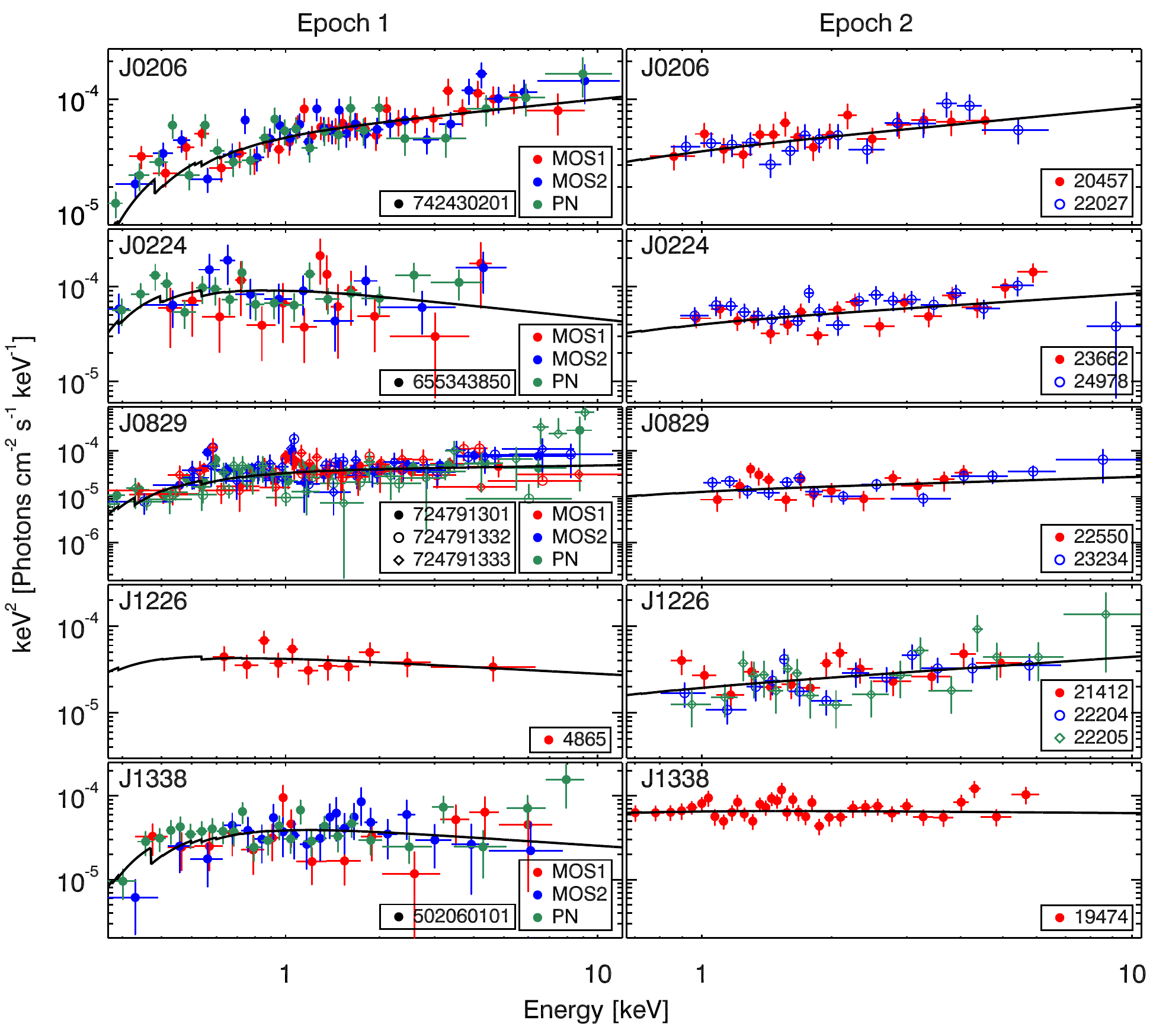}
\caption{X-ray spectra of the five CLQs. 
For each CLQ, the left panel is Epoch1, and the right panel is Epoch 2. The black lines are the best-fit models. Different symbols are from different observations. The first four CLQs became fainter in Epoch 2, consistent with the optical state. The last one, J1338 re-brightened quickly near the epoch of the {\it Chandra} follow-up (confirmed by the Magellan optical spectroscopic follow-up).
No strong intrinsic absorption is detected in the faint state.
Changes in the power-law continuum were detected, both in the photon indices and the fluxes.
\label{fig:Xray_spec}}
\end{figure*}

\begingroup
\renewcommand*{\arraystretch}{1.2}
\begin{deluxetable*}{crrrrrcr}
\tablecaption{Changing-obscuration Model \label{tab:Xray_varing}}
\tablewidth{1pt}
\tablehead{
\colhead{Name} &
\colhead{$\chi^2$} &
\colhead{dof} &
\colhead{$\chi^2_{\rm red}$} &
\colhead{$\chi^2_{\rm abs}$} &
\colhead{dof$_{\rm abs}$} &
\colhead{$\chi^2_{\rm red, obsc}$} &
\colhead{${\rm N_{H, obsc}}$} \\
 & & & & & & & $10^{22}~{\rm cm}^{-2}$
}
\startdata
J0206 & 22.653 & 29 & 0.78 & 50.394 & 31 & 1.63 & 0.42$_{-0.09}^{+0.10}$ \\
J0224 & 54.727 & 35 & 1.56 & 107.572 & 37 & 2.91 & 1.15$_{-0.11}^{+0.12}$ \\
J0829 & 45.523 & 27 & 1.69 & 253.722 & 29 & 8.75 & 2.61$_{-0.29}^{+0.29}$ \\
J1226 & 61.520 & 46 & 1.34 & 114.350 & 48 & 2.38 & 1.17$_{-0.17}^{+0.16}$ \\
\enddata
\tablecomments{
$\chi^2_{\rm red,obsc}$, the reduced $\chi^2$ values of the changing-obscuration model, is larger than $\chi^2_{red}$ of the independent best-fit model from Table~\ref{tab:Xray_model}.
}
\end{deluxetable*}
\endgroup

\begin{figure*}[!ht]
\centering
\includegraphics[width=0.97\textwidth]{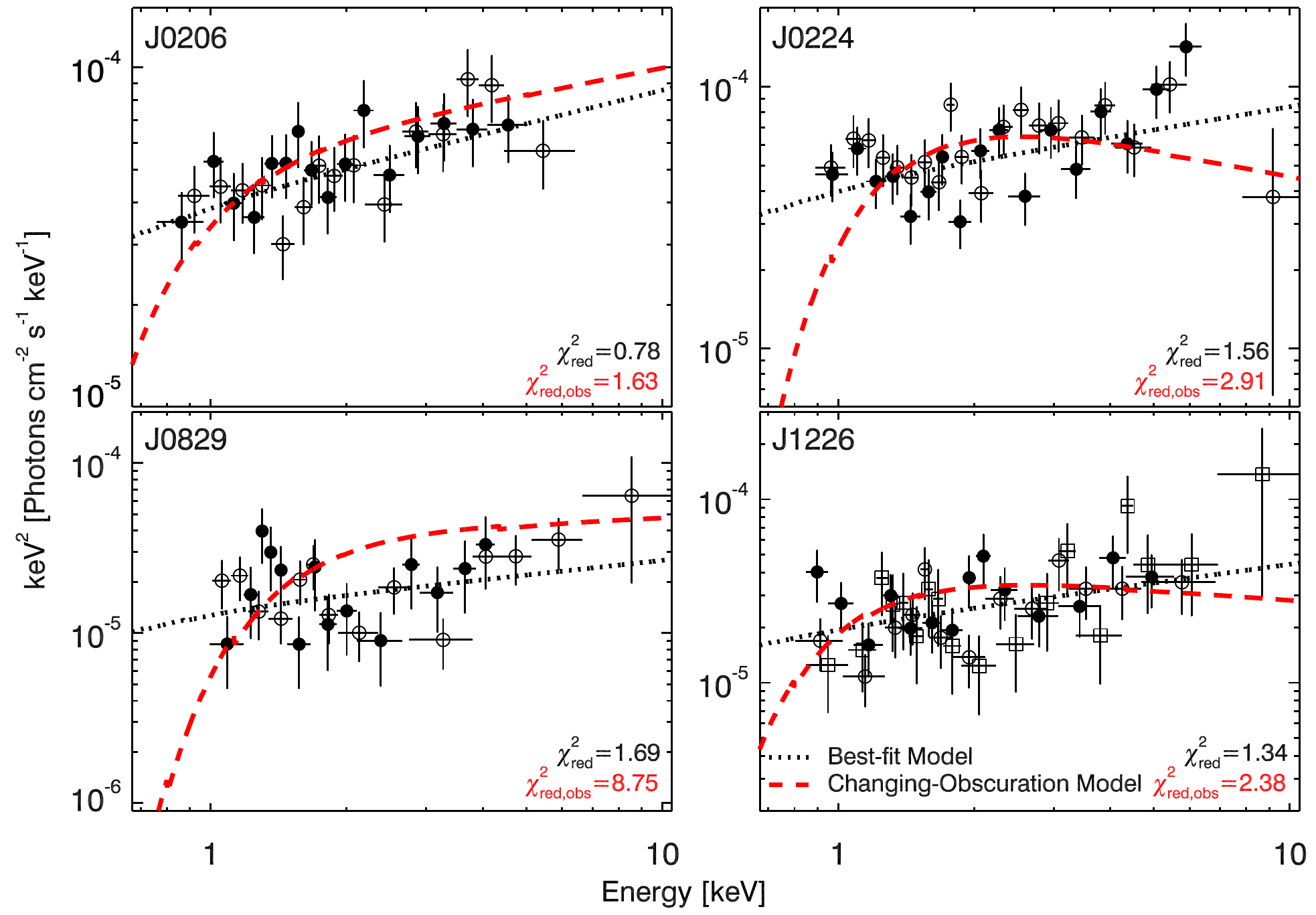}
\caption{
X-ray spectral fitting for four CLQs in our {\it Chandra} X-ray faint-state observations (J1338 is not included). The different symbols refer to {\it Chandra} observations with different ObsID (filled circle, open circle, and open square with observation time from early to late). 
The black dotted lines are the best-fit model to Epoch 2 (faint-state) X-ray spectra.  
The red dashed lines are the changing-obscuration model fits to the Epoch 2  spectra, assuming Epoch 2 and Epoch 1 intrinsic spectra are identical with varying obscuration.
The X-ray spectra are more consistent with intrinsic variation (both in $\Gamma_{X}$ and fluxes), rather than the changing-obscuration model.
\label{fig:Xray_varying}}
\end{figure*}

Using the bright state spectra, with their prominent broad H$\beta$ emission, we find the virial $M_{\rm BH}$ based on Equation \ref{eq:MBH} for these CLQs are $10^{8-9}~M_{\odot}$ (see individual results in Table \ref{tab:Optical_prop}). Our $M_{\rm BH}$ measurements are consistent with the results in \citet{Wu2022}, though slightly lower ($0.04-0.29$ dex), since we accounted for the host galaxy contribution, thus yielding a lower AGN continuum emission.

Figure \ref{fig:LC} and \ref{fig:LC_one} show the multiwavelength light curves of the five new CLQs, from top to bottom panels, in X-ray, optical $g$ and $r$ bands, and MIR $W1$ and $W2$ bands. In the optical, the photometric data have been calibrated to the SDSS filter magnitude (as described in \textsection \ref{sec:optical_imaging}). The multiwavelength light curves show highly coordinated variability trends across the different wavebands. 

Typical quasar variability is stochastic. In contrast, J1226 faded quite steadily for more than two decades from 1999 to 2021 in all bands from X-ray to MIR. 
The light curves show that J1226 is different from the other four CLQs, which have shorter timescale stochastic variability. Recent ZTF photometric data show signs of re-brightening in all five CLQs. 

The {\it Chandra} observation shows J1338 to be brighter than its previous X-ray observation (described in \textsection \ref{sec:Results_Xray}). It brightened by 0.7 mag in the $g$ band and 0.3 mag in the $r$ band within two months from our MMT observation to the {\it Chandra} observation, according to our photometric monitoring using the 1.2m FLWO telescope. The $r$-band photometry close to the {\it Chandra} observation is actually 0.1 mag brighter than the one close to the {\it XMM-Newton} observation. In the Magellan spectrum, 11 months after the MMT observation, the broad H$\beta$ line reappeared. Therefore, we remove J1338 in the analysis in Section \ref{sec:Results_Xray} as our {\it Chandra} X-ray observation was not performed in its faint state.

\subsection{X-ray Results} \label{sec:Results_Xray}
Since the primary physical puzzles concern the SMBH accretion rate and intrinsic absorption, X-ray observations are the ideal probes. 
We firstly fit the X-ray spectra in both states independently. Figure \ref{fig:Xray_spec} shows the X-ray spectral fitting results for the five CLQs in Epoch 1 (left panels) and Epoch 2 (right panel), with the model described in \textsection \ref{sec:xray_fit}. 
Within each panel, the observations taken for the same object with the same telescope within one week are fit together. 
The different colors/symbols are data from different observations (or cameras). We summarize the X-ray fitting results in Table \ref{tab:Xray_model}. 

No strong intrinsic absorption is detected in either the bright or faint states for these quasars. The faint-state X-ray spectra are well fit with zero intrinsic column density, ${\rm N_{H, intrin}}$, with $1\sigma$ upper limits smaller than or comparable with the ${\rm N_{H, intrin}}$ values in the bright-state X-ray spectra. This finding is consistent with previous X-ray studies of CLQs \citep{LaMassa2015, Ai2020} and in CL AGN such as Mrk 590 \citep{Denney2014} and Mrk 1018 \citep{Husemann2016}. 

Instead, we detected changes in the power-law continuum, both in photon index and flux. Apart from J1338, which re-brightened rapidly, in the other four CLQs, the intrinsic fluxes in Epoch 2 are lower than those in Epoch 1. As shown in Figure \ref{fig:LC}, their X-ray emission generally changes together with the optical variability. The X-ray spectra are harder in the faint state, consistent with the trend in ensembles of AGN \citep[e.g.,][]{Dong2014}. As the energy range of {\it Chandra} (0.5 - 7 keV) is different from {\it XMM-Newton} (0.3 - 12 keV), we tested the spectral shape changes using the same energy range of 0.5 - 7 keV.  Restricting the  {\it XMM-Newton} fitting to the 0.5 - 7 keV energy range, the $\Gamma_{X}$ values differ by $\lesssim0.02$, which is much smaller than the typical uncertainty of $\Gamma_{X}$. Therefore, the conclusion of the X-ray spectra change is not affected by the  difference between the two energy ranges.

As X-ray data are sensitive to absorption, we use the X-ray spectra to test the changing-obscuration model. Assuming varying absorption is the cause of the changes, we fix the AGN power-law continuum parameters in the faint state to the measurements obtained in the bright state, with only quasar intrinsic column density as a free parameter. Figure \ref{fig:Xray_varying} shows the results for the four CLQs observed in the faint state by {\it Chandra}. The plotted data are from {\it Chandra} in the faint state. The dotted black line is the independent best-fit model in the faint state. The dashed red line is the best-fit changing-obscuration model to the faint-state data. The results demonstrate that the changing-obscuration model does not match the observations. The reduced $\chi^2$ value, $\chi^2_{\rm red, obsc}$ of this changing-obscuration model is larger than the results of the independent best-fit model (summarized in Table \ref{tab:Xray_varing}). For these CLQs, the X-ray data are not consistent with a varying absorption scenario.

\begin{figure*}[!ht]
\centering
\includegraphics[width=1.0\textwidth]{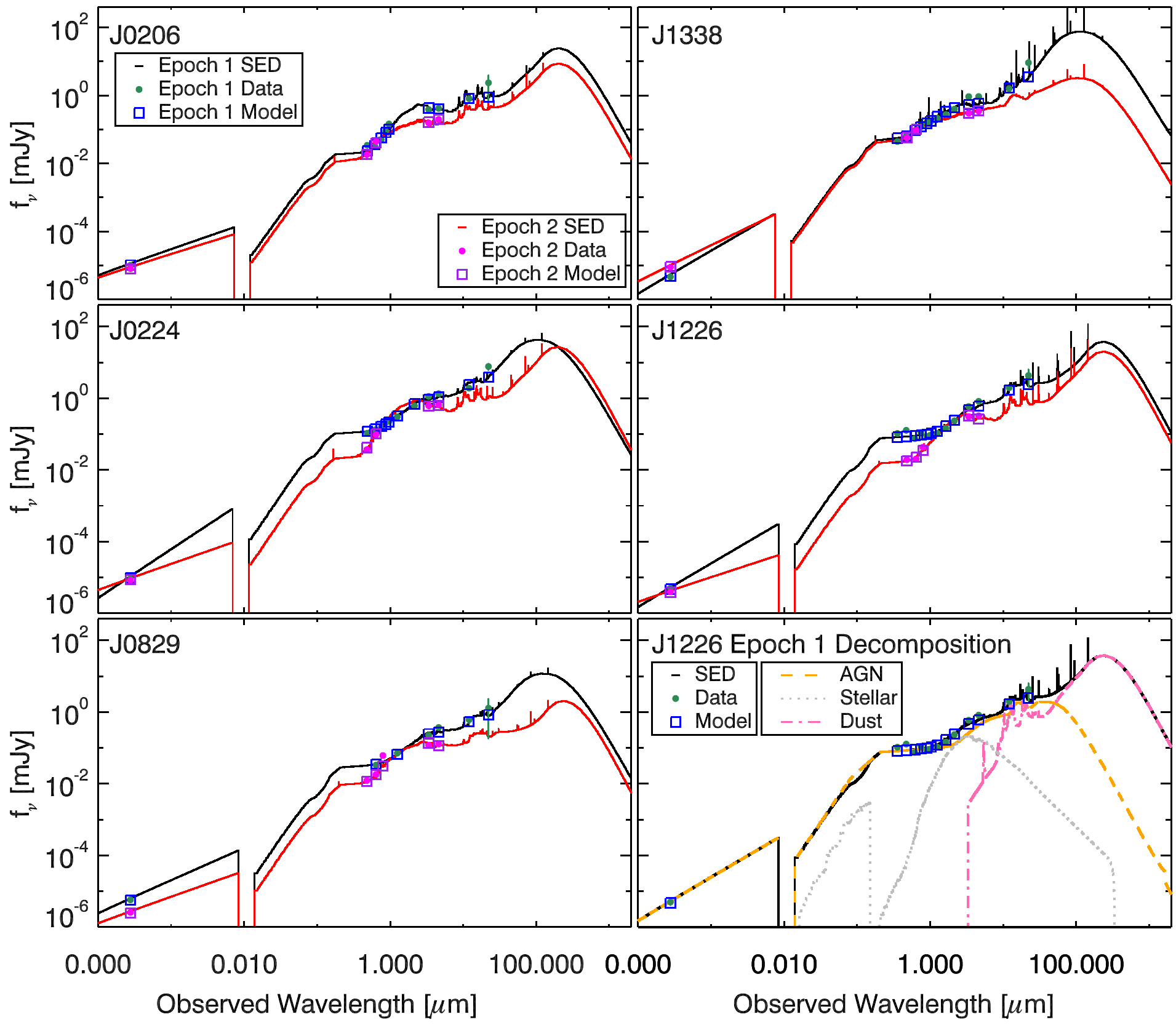}
\caption{
SED fitting to the five CLQs. The solid lines are the best-fit SED models in Epoch 1 (black) and Epoch 2 (red). The solid circles are imaging observation data in Epoch 1 (green) and Epoch 2 (magenta), as well as model fluxes in corresponding bands in Epoch 1 (blue) and Epoch 2 (purple). The bottom right panel shows an example of the SED decomposition of J1226 in Epoch 1 in consists of components including AGN (orange; dashed), host galaxy stellar emission with extinction attenuated (gray; dotted), and dusty torus (pink; dash-dotted). 
There is stronger variability in the X-ray, UV/optical, and mid-infrared than in the near-infrared, which is dominated by  host galaxy emission. The dramatic variability in the MIR and the trend of larger variability in the longer wavelengths are not consistent with the changing-obscuration model.
\label{fig:SED}}
\end{figure*}

\subsection{CLQ SED Variability}
The CLQs show multiwavelength variability.
Figure \ref{fig:SED} illustrates the comparison between the SEDs in Epoch 1 (black) and Epoch 2 (red). We fit the SEDs with components including an AGN (X-ray and UV to IR), host galaxy, and dusty torus (as described in \textsection \ref{sec:SED_fit}). The right bottom panel shows an example of the SED decomposition for J1226 in the bright state. In four CLQs, J0206, J0224, J0829, and J1226, Epoch 2 is fainter than Epoch 1 throughout the electromagnetic spectrum. We integrate the best-fit SED model to obtain the bolometric luminosity, $L_{\rm bol}$, and use the virial $M_{\rm BH}$ obtained from the optical spectra to calculate the Eddington ratio, $\lambda_{\rm Edd}$. Only for J1338 is the $L_{\rm bol}$ higher in Epoch 2 than in Epoch 1. The $L_{\rm bol}$ values span $10^{44.7-45.6}$ erg s$^{-1}$ and the $\lambda_{\rm Edd}$ values range from $10^{-2.2}$ to $10^{-0.5}$ (0.6\% to $\sim$30\%; summarized in Table \ref{tab:SED_results}). We also measure the UV luminosity ($\lambda L_{2500\AAm}$) and UV to X-ray slope ($\alpha_{\rm OX}$) from the SED fitting.
Use of the SED fitting is more accurate for this purpose than simply extrapolating rest frame luminosities from total observed fluxes, because we are interested in changes to the AGN emission, and our SED fitting models the host galaxy and AGN components separately, especially important for the faint states.

The entire SED variability shape is not consistent with varying obscuration; while the mid-infrared flux varies significantly, it is at most weakly affected by dust extinction. The extinction coefficient in the $g$ band is a factor $>15$ greater than that in the $W1$ band, according the extinction law of \citet{Fitzpatrick1999}, across a range $R_V$ values (see Table \ref{tab:Extinction}). All five CLQs vary by more than 0.5 mag in $W1$ band, which cannot be caused by changing extinction given the corresponding $\sim  1-2 $ mag optical variability. Some of the CLQs show larger variability in the $W2$ band than in the $W1$ band, also inconsistent with an extinction model.

It is more likely the changes in these CLQs are due to a changing accretion state of the SMBH, whereby the multiwavelength emission varies accordingly. The accretion rate of the central SMBH decreased, so the whole system dimmed. The AGN emission throughout the entire electromagnetic spectrum -- the UV-optical emission from the accretion disk, the X-ray emission from the corona, and the MIR emission from the re-radiation in the dusty torus -- decreased as the luminosity from the central engine decreased. As a consequence, the emission from the BLR, photoionized by the UV emission from the accretion disk, fades. The SEDs show stronger variability in the X-ray, UV optical, and MIR, than in the near-infrared. This is also reasonable in this scenario, because the emission in the NIR is dominated by the host galaxy, which does not vary. If the common model of AGN structure with hotter regions at smaller radii is correct, then densely cadenced multiwavelength photometry throughout a state transition should verify that continuum changes generally propagate with time from shorter to longer wavelengths.

\begin{deluxetable*}{cccccc}
\tablecaption{SED Fitting Results \label{tab:SED_results}}
\tablewidth{1pt}
\tablehead{
\colhead{Name} &
\colhead{State} &
\colhead{log($L_{\rm bol}$)} &
\colhead{log($\lambda_{\rm Edd}$)} &
\colhead{log($\lambda L_{2500\AAm}$)} &
\colhead{$\alpha_{\rm OX}$} \\
 & & erg s$^{-1}$ & & erg s$^{-1}$ &
}
\startdata
J0206 & 1 & 44.94$\pm$0.03 & -2.06$\pm$0.06 & 44.02$\pm$0.06 & 1.10$\pm$0.03 \\
J0206 & 2 & 44.81$\pm$0.03 & -2.19$\pm$0.06 & 43.80$\pm$0.09 & 1.07$\pm$0.03 \\
\hline
J0224 & 1 & 45.26$\pm$0.05 & -1.35$\pm$0.07 & 44.63$\pm$0.07 & 1.29$\pm$0.03 \\
J0224 & 2 & 44.78$\pm$0.03 & -1.83$\pm$0.06 & 43.93$\pm$0.10 & 1.16$\pm$0.04 \\
\hline
J0829 & 1 & 45.32$\pm$0.04 & -1.04$\pm$0.17 & 44.60$\pm$0.06 & 1.22$\pm$0.02 \\
J0829 & 2 & 44.90$\pm$0.02 & -1.46$\pm$0.17 & 44.11$\pm$0.01 & 1.21$\pm$0.02 \\
\hline
J1224 & 1 & 45.64$\pm$0.01 & -0.59$\pm$0.15 & 45.03$\pm$0.01 & 1.35$\pm$0.03 \\
J1224 & 2 & 45.11$\pm$0.02 & -1.11$\pm$0.15 & 44.33$\pm$0.03 & 1.23$\pm$0.02 \\
\hline
J1338 & 1 & 45.15$\pm$0.03 & -1.29$\pm$0.13 & 44.51$\pm$0.03 & 1.28$\pm$0.02 \\
J1338 & 2 & 45.17$\pm$0.05 & -1.27$\pm$0.14 & 44.44$\pm$0.08 & 1.19$\pm$0.03 \\
\enddata
\end{deluxetable*}

\begin{figure*}
\centering
\includegraphics[width=0.85\textwidth]{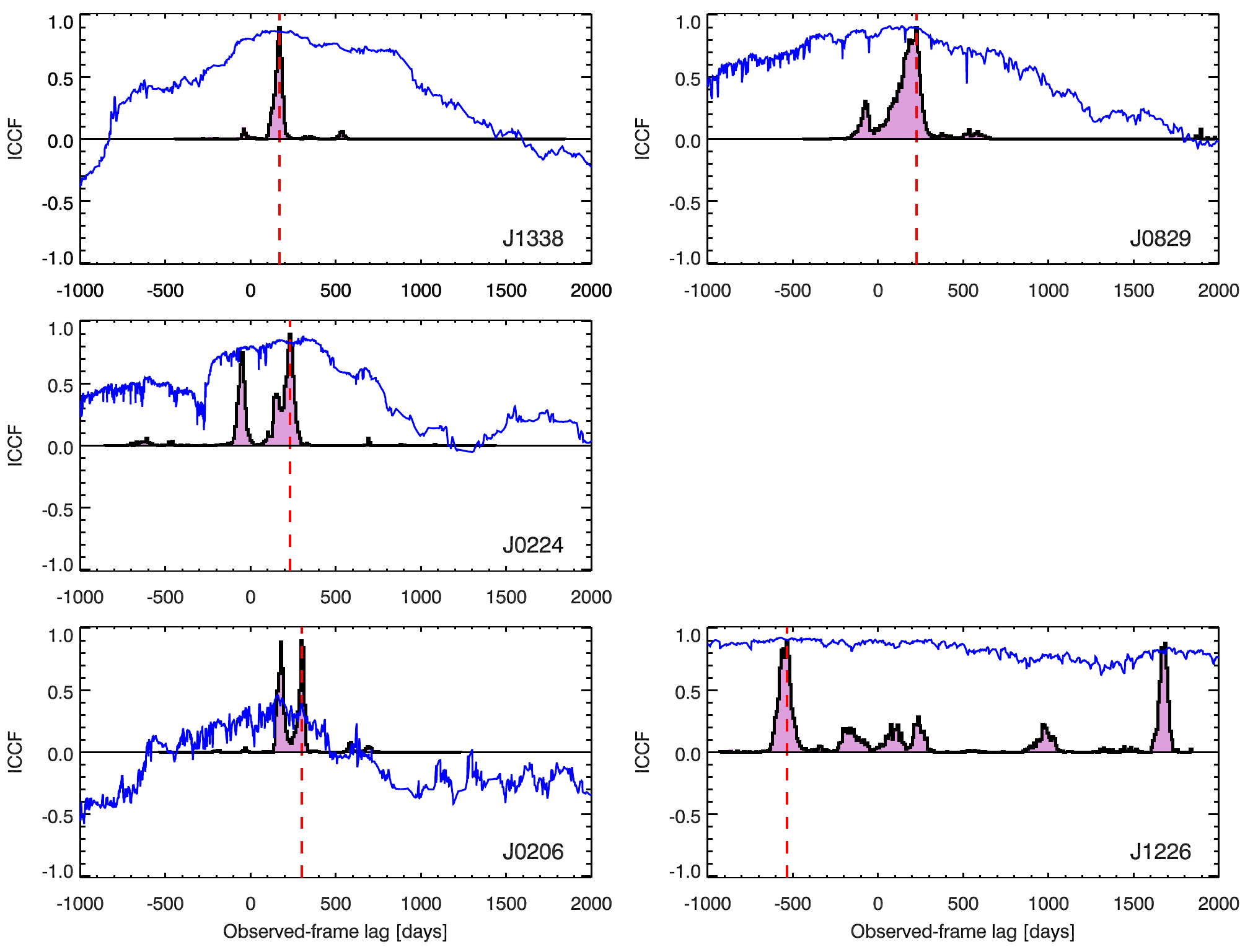}
\caption{
Time lag measurements between the WISE $W1$ band and optical $r$ band light curves. The blue solid line is the ICCF. The purple histogram (with black solid outline) shows the \javelin posterior lag distribution in observed frame. The vertical red dashed line shows the primary peak of the JAVELIN posterior distribution. The top two panels show the two CLQs with high-fidelity lag measurements, J1338 (left) and J0829 (right). The middle panel shows one CLQ, J0224, with a marginal lag measurement, with one positive primary peak and one negative lag with lower probability. In the bottom panels, either the \javelin or ICCF algorithm fails to measure a robust MIR lag for J0206 (low ICCF value) and J1226 (many aliases).
\label{fig:DustRM}}
\end{figure*}

\begin{figure}
\centering
\includegraphics[width=0.48\textwidth]{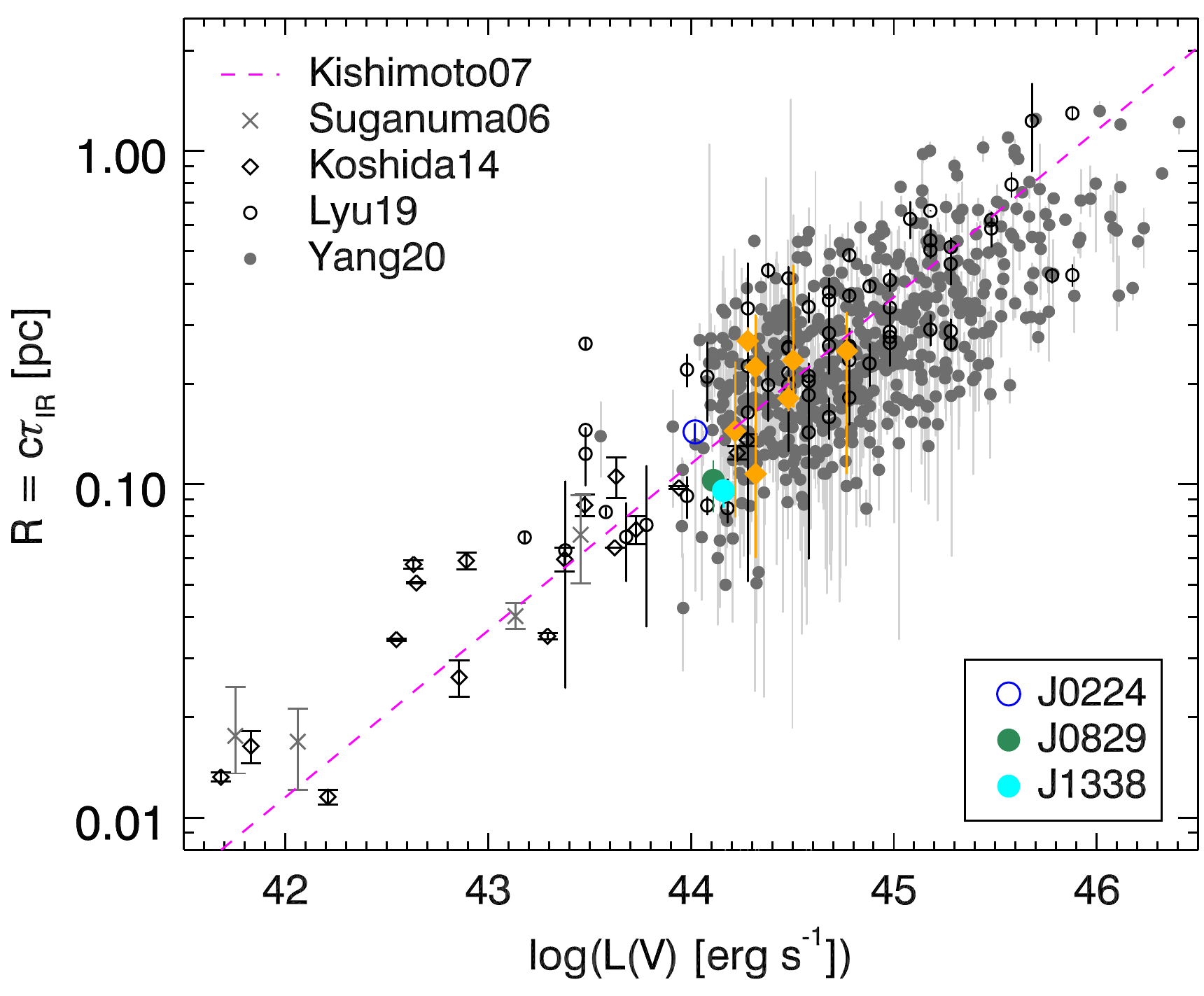}
\caption{
The dusty torus size, inferred from the rest-frame IR lag, vs. the rest-frame optical quasar luminosity. The black diamonds and gray crosses are lag measurements between $K$ band and optical band for nearby AGN \citep{Suganuma2006, Koshida2014}. The black open circles and gray solid dots are lag measurements between WISE $W1$ band and optical band for PG quasars \citep{Lyu2019} and SDSS quasars \citep{Yang2020}. The orange solid diamonds are lag measurements in \citet{Yang2020} for CLQs in literature. The blue (open), green (solid), and cyan (solid) are the lag measurements for CLQs in this work for J0224, J0829, and J1338, respectively. The magenta line is the R−L relation in \citet{Kishimoto2007}. 
\label{fig:RL}}
\end{figure}

\subsection{Dust Echoes in CLQs}
If the MIR light is reprocessed light from the torus, its echo of the UV/optical continuum variations measures the average light-crossing time (hence a typical size) of the dust torus to the central engine. 
We measure the time lags between the MIR $W1$-band and optical $r$-band light curves of these CLQs. We use the public code \javelin, which fits the light curves using a damped random walk (DRW) model and aligns them to recover the time lag \citep{Zu2011}. The DRW model has proven to be a reasonably good prescription to describe the optical continuum variability of quasars \citep[e.g.,][]{Kelly2009, MacLeod2010}.
The inner boundary of the torus is thought to be set by dust sublimation, roughly at $\sim1500\,$K, thus the innermost region is mainly emitting in the NIR \citep[e.g.,][]{Kishimoto2007}.  For the five CLQs with redshift at 0.36-0.64, WISE $W1$ (3.4 $\mu$m) band probes the dust emission at rest frame 2.1-2.5 $\mu$m.
MIR $W1$-band emission has been used previously to measure the torus radii for quasars \citep[e.g.,][]{Lyu2019, Yang2020}, while in nearby AGN, NIR $K$-band emission has been used \citep[e.g.,][]{Suganuma2006, Koshida2014}.

Robustly measuring the lag between two sets of light curves depends on the quality of the light curves (e.g., duration, cadence, signal-to-noise ratio), as well as their intrinsic variability \citep{Yang2020}. To explore all possible lags while preserving significant overlaps between MIR and optical light curves, we allow a large lag search window of [-1000, 2000] days. In figure \ref{fig:DustRM}, we show the lag measurement results for the five CLQs. We measure high-fidelity lags (with one primary lag peak) using \javelin for J1338 ($166_{-19}^{+12}$ days) 
and J0829 ($200_{-39}^{+29}$ days), consistent with the peak from the interpolated cross-correlation function \citep[ICCF;][]{Gaskell1987, Peterson1998}. For J0224, we obtained a positive lag of $232_{-15}^{+14}$ days and a negative lag with lower probability. For J0206, the ICCF value is small ($<0.5$) and \javelin obtains two equal positive peaks at 179 and 299 days. J1226 faded all the way over two decades, so it is hard to measure to a time lag between the two light curves without an extreme value. As a consequence, as shown in the right bottom panel in Figure \ref{fig:DustRM}, the ICCF value of J1226 is large across the whole lag search window and \javelin\, identifies too many alias peaks.

Figure \ref{fig:RL} shows the correlation between the dusty torus size, inferred from the rest-frame IR lag, and the rest-frame optical quasar luminosity\footnote{To directly compare with earlier studies, following \citet{Yang2020}, we convert the bolometric luminosity to $V$-band luminosity (assuming a bolometric correction of 10 in $V$ band).}. \citet{Yang2020} measured a high-fidelity lag between MIR and optical for a sample of 587 quasars at $\left<z\right>\sim 0.8$. Among these quasars, seven quasars were reported as CLQs (shown as orange solid diamonds), including 4 in \citet[][J0023+0035, J2146+0009, J2252+0109, J2333-0023]{MacLeod2016}, 1 in \citet[][J2317+0114]{MacLeod2019}, 1 in \citet[][J2343+0038]{Yang2020EVQ}, and 1 in \citet[][J0212-0030]{Green2022}. The dusty torus size of the two CLQs with high-fidelity lags (J0829 and J1338) and one CLQ J0224 (using the primary positive lag), as well as the seven CLQs in literature, are fully consistent with the torus $R-L$ relation \citep{Kishimoto2007}, including the torus size measurements for nearby AGN \citep{Suganuma2006, Koshida2014} and normal quasars \citep{Lyu2019, Yang2020}. The MIR variability echoes the variability in the optical, with a time lag expected from the light-crossing time of the dusty torus.

\begin{figure}
\centering
\includegraphics[width=0.48\textwidth]{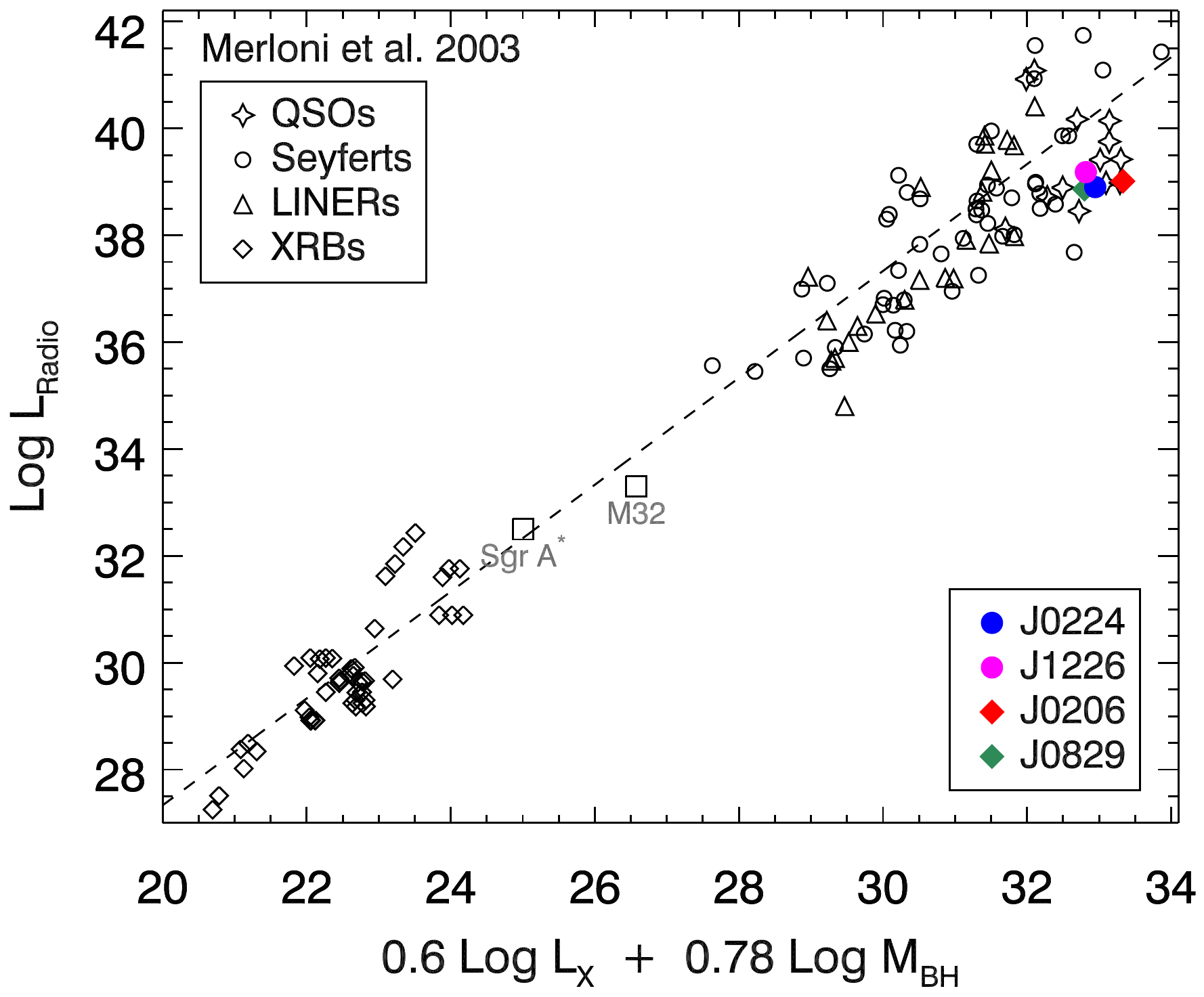}
\caption{
Edge-on view of the fundamental plane of black hole activity. The sources in black are from \citep{Merloni2003}, including QSOs (stars), Seyferts (circles), low-luminosity AGN in LINERs (triangles), XRBs (diamonds), and Sgr A$^*$ and M32 (squares). The colored dots are the four CLQs observed by VLA in the faint state, with filled circles denoting sources with unresolved radio emission, and filled diamonds denoting extended radio emission.  
\label{fig:FP}}
\end{figure}

\begin{figure}
\centering
\includegraphics[width=0.47\textwidth]{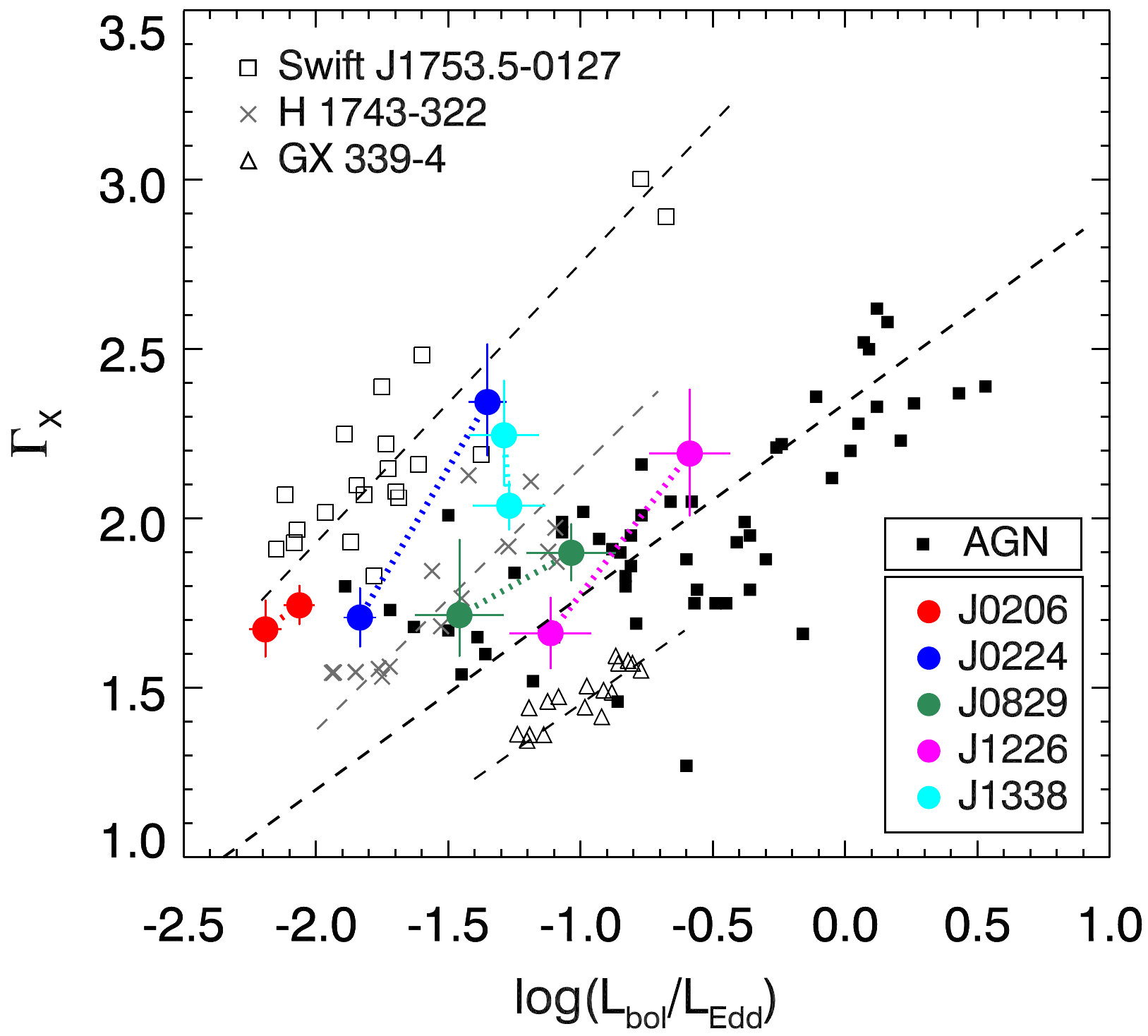}
\caption{
Relation between X-ray photon index, $\Gamma_X$, and Eddington ratio (${\rm L_{bol}/L_{Edd}}$) in logarithm, for AGN and XRBs. The color dots are our five CLQs, including J0206 (red), J0224 (blue), J0829 (green), J1226 (magenta), and J1338 (cyan). The dotted lines connect the CLQ in bright state and faint state. The black squres are measurements from AGN in \citet{Dong2014}. There are some individual XRBs, including Swift J1753.5$-$0127 (open square), H1743$-$322 (cross), and GX 339$-$4 (open triangle). One can see the same trend, softer when brighter, in AGN and XRBs. The individual CLQs follow this trend well.  J1338 was unfortunately not captured in its faint state during our {\it Chandra} observation.
\label{fig:Gamma}}
\end{figure}

\begin{figure}
\centering
\includegraphics[width=0.47\textwidth]{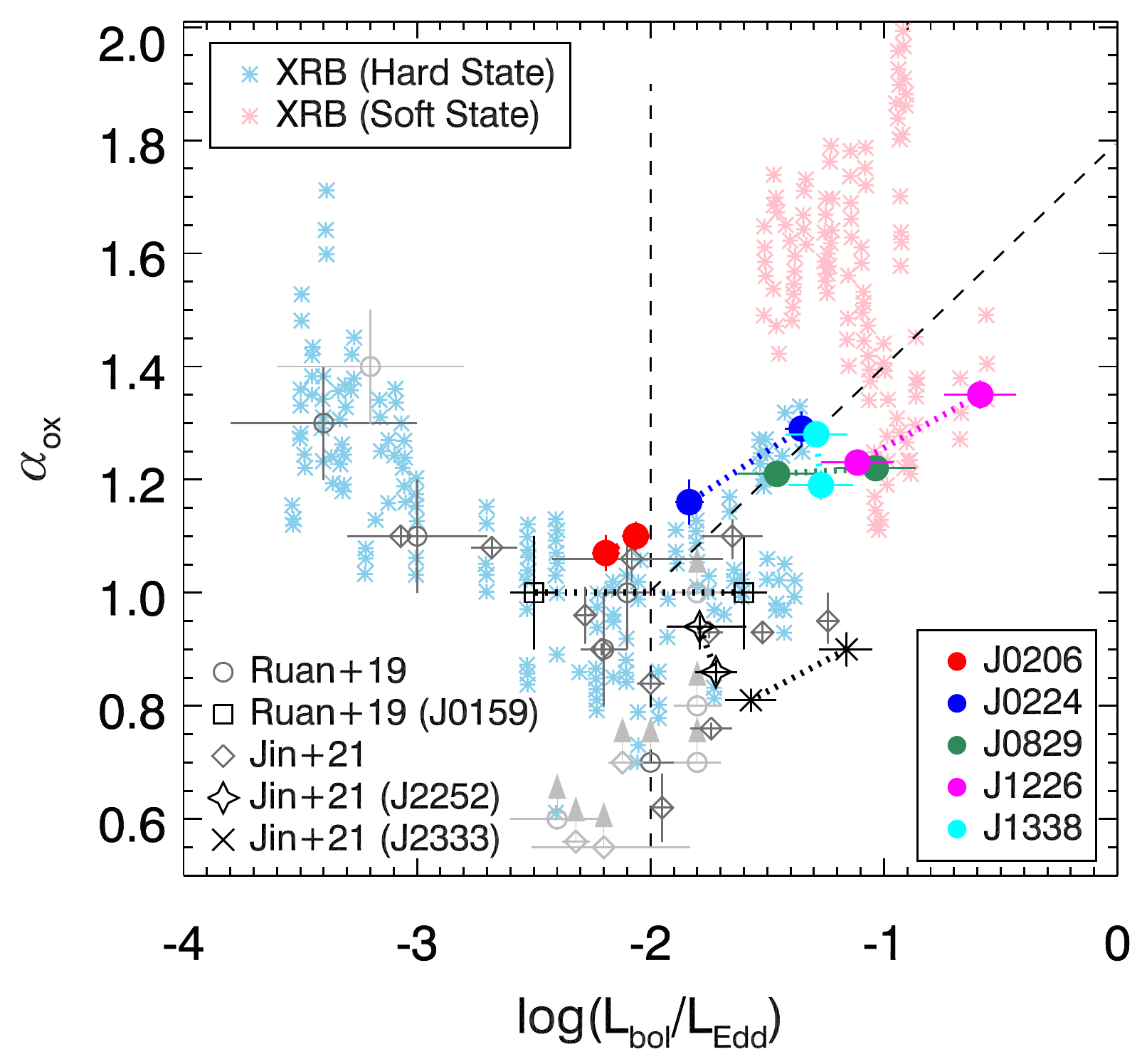}
\caption{
Relation between the UV–to–X-ray spectral index ($\alpha_{\rm OX}$) and the logarithm of the Eddington ratio log(${\rm L_{bol}/L_{Edd}}$), for AGN and XRBs. 
Both $\alpha_{\rm OX}$ and $L_{bol}$ are taken from our SED fitting.
Colored solid dots are our 5 {\it Chandra} CLQs as in Figure~\ref{fig:Gamma}. The gray symbols are from CLQs in the faint states (open circles) and lower limits in the bright states (open circle with arrows) in \citet{Ruan2019}, and CLQs in the faint states (open diamonds) and lower limits in the bright states (open diamonds with arrows) in \citet{Jin2021}. The black symbols are CLQs with X-ray observations in both states for J0159 (squares) analyzed in \citet{Ruan2019} and J2252 (star) and J2333 (cross) in \citet{Jin2021}. 
The dotted line connects one CLQ observed in two states.
The asterisks show results from the Galactic XRB GRO J1655-40 \citet{Sobolewska2011} in hard state (sky blue) and soft state (pink), extrapolated via mass scaling to Type 1 AGN (see their Figure 4). 
\label{fig:aox}}
\end{figure}

\section{Discussion}
\subsection{Fundamental Plane of Black Hole Activity}
Accretion onto BHs illuminates important physics from stellar mass BHs in XRBs to SMBHs. The fundamental plane of black holes shows the correlation of radio luminosity, X-ray luminosity, and black hole mass for hard-state XRBs and their supermassive analogs \citep[e.g.,][]{Merloni2003, Falcke2004, Plotkin2012, Dong2014, Gultekin2019}. The theoretical framework behind the fundamental plane typically assumes a black hole fed by a RIAF that is launching a compact, partially self-absorbed synchrotron jet (implying unresolved radio emission with a flat/inverted spectrum; e.g., \citealt{Heinz2003}). 

Using the VLA radio observations, we obtain the radio luminosity at 5 GHz in the rest frame, $L_{\rm 5GHz}$, typically $\sim10^{39}$ erg s$^{-1}$ for the four CLQs (summarized in Table \ref{tab:Radio_obs}). In Figure \ref{fig:FP}, we plot the CLQs on the edge-on view of the fundamental plane in \citet{Merloni2003}. 
The sources in black are QSOs \citep[stars;][]{Kaspi2000}, Seyferts (circles), low-luminosity AGN in LINERs (triangles), XRBs \citep[diamonds;][see references therein]{Merloni2003}, and Sgr A$^*$ and M32 \citep[squares;][]{Schodel2002,Verolme2002}. The correlation extends over many decades in black hole mass and luminosity, including many different sources, from stellar mass BHs to SMBHs. The CLQs appear to follow the fundamental plane relation. However, there are several caveats to bear in mind that complicate physical interpretations with current data.  Foremost, although we detect radio emission from all four CLQ targets in their faint states, we stress that it is unclear if this emission is related to the activation of a compact jet, or if the radio emission is instead a relic from the previous QSO phase.  It is also unknown if our targets' radio emission have flat/inverted spectra, and only two of our targets remain unresolved at VLA spatial resolutions (J0224 and J1226). 
 We use different symbols in Figure \ref{fig:FP} to distinguish point source (filled circle) and possible extended (filled diamond) sources. 

\subsection{Analogy between AGN and XRBs}
Accretion state transitions have been detected in many XRBs. The X-ray power-law continuum slope correlates strongly with Eddington ratio, $\lambda_{\rm Edd}$ in individual XRBs as their accretion state varies, such as Swift J1753.5$-$0127 \citep{Cadolle2007}, H1743$-$322 \citep{Jonker2010}, and GX 339$-$4 \citep[e.g.,][]{Corbel2013}. Similar trends are seen (so far) across samples of quasars and Seyferts \citep[e.g.,][]{Trichas2013, Brightman2013}, but have rarely been tracked in individual AGN, much less luminous quasars. 
Figure \ref{fig:Gamma} shows the X-ray power-law slope, $\Gamma_{\rm X}$, vs. the Eddington ratio. The colored dots are our five CLQs; dots of the same color represent the same object in different states. The black filled squares are individual AGN in the sample of \citet{Dong2014}, and the other black symbols are individual XRBs. Each individual CLQ follows the same softer-when-brighter trend, probably due to changes in the hot corona temperature or opacity \citep[e.g.,][]{Petrucci2018, Middei2019}

There is variability in both the UV/optical and X-ray. The parameter $\alpha_{\rm OX}$ provides a measurement of the ratio of disk to corona emission. Large $\alpha_{\rm OX}$ corresponds to strong UV relative to X-rays (i.e., a soft SED). Figure \ref{fig:aox} shows the behavior of $\alpha_{\rm OX}$ vs. Eddington ratio, $\lambda_{\rm Edd}$. The gray symbols are CLQ measurements in the faint state \citep{Ruan2019, Jin2021} and (ROSAT) X-ray upper limits in the bright state (so lower limits in $\alpha_{\rm OX}$, shown with upward arrows). The black symbols are three known CLQs with X-ray measurements in both states, including J0159 \citep{Ruan2019}, and J2252 and J2333 \citep{Jin2021}. 

There is some evidence that $\alpha_{\rm OX}$ varies with $\lambda_{\rm Edd}$ in two branches around a critical Eddington ratio of $\sim10^{-2}$ \citep[1\%; e.g.,][]{Ruan2019, Jin2021}. When $\lambda_{\rm Edd} \gtrsim 1\%$, as is the case for our CLQ observations, $\alpha_{\rm OX}$ hardens (decreases) when $\lambda_{\rm Edd}$ decreases.  By contrast, when $\lambda_{\rm Edd} \lesssim 1\%$, $\alpha_{\rm OX}$ softens (increases) when $\lambda_{\rm Edd}$ is smaller. While there is evidence to support this trend from ensembles of AGN each observed once, it is key to track the trends among individual AGN observed at different accretion rates because other factors - most importantly SMBH mass and inclination angle - are obviously fixed across epochs for individual AGN.
Our CLQs are mostly $\lambda_{\rm Edd} \gtrsim 1\%$, so the individual CLQs vary such that $\alpha_{\rm OX}$ increases (softens) with  $\lambda_{\rm Edd}$. J0206, with $\lambda_{\rm Edd}$ slightly lower than $10^{-2}$, also follows this trend. J1338 is an outlier, but as we mentioned it varies quickly, and we did not catch the desired faint state in X-rays. For J0159, with one epoch at $\lambda_{\rm Edd} > 10^{-2}$ and another epoch at $\lambda_{\rm Edd} < 10^{-2}$,  \citet{LaMassa2015} reported no detected 
change in X-ray slope or (albeit with large errors) in $\alpha_{\rm OX}$ .

In Figure \ref{fig:aox}, the asterisks are predictions for AGN from \citep{Sobolewska2011}, based on modeling the observed X-ray spectral evolution of the XRB GRO J1655$–$40 during its accretion state transition and scaling the SED evolution up to AGN SMBH masses. The predicted data from XRBs also show a critical Eddington ratio of about 1\%. 
The analogy of SMBHs and stellar mass black holes motivates a theoretical model to explain the phenomena. 
Referring to theoretical models of XRBs, \citep{Ruan2019} suggested that 
at high Eddington ratios, accretion proceeds via a geometrically thin accretion disk \citep{Shakura1973}. As $\lambda_{\rm Edd}$ decreases, there is less UV emission from the thin disk with lower apparent temperature, so $\alpha_{\rm OX}$ decreases. 
When the $\lambda_{\rm Edd}$ is lower than 1\%, a jet may be launched or the inner region of the thin disk may progressively evaporate into a hot radiatively inefficient accretion flow (RIAF) that is possibly advection-dominated \citep{Narayan1994}. 
Thus, $\alpha_{\rm OX}$ increases again possibly due to the emergence of UV emission from either a jet or an advection-dominated accretion flow (ADAF). For our CLQs, the $\lambda_{\rm Edd}$ is not far below $10^{-2}$, and the radio loudness is $\sim10^{-5}$ (see Table \ref{tab:Radio_obs}), indicating no powerful jet launch, so our observations are consistent with the models.  Multi-epoch observations of CLQs at even lower Eddington ratios are warranted.  
The analogy with XRBs may break down at very low $\lambda_{\rm Edd}$;
weak AGN in LINERs were found with $\lambda_{\rm Edd} < 10^{-4}$ and $\alpha_{\rm OX} \approx 1\pm0.1$ \citep{Eracleous2010}. 
However, evidence so far indicates that most AGN with $\lambda_{\rm Edd}< 1\%$ will not show broad emission lines, so the necessary observations may require following significant intrinsic (i.e., not absorption-related) variability in type 2 AGN. In highly obscured type 2 AGN, even when there is strong variability in the MIR ($>3$ mag), the optical continuum emission could be highly obscured preventing significant optical ($<0.2$ mag) or broad emission line variability \citep{Yang2019}. Strong  variability in ``naked type 2" AGN is relatively rare, but worth pursuing; they may on closer inspection be found to have weak broad H$\beta$ emission, and/or broad H$\alpha$ \citep[e.g., associated with types 1.8 or 1.9;][]{Barth2014, Lopez-Navas2022}. 

\section{Summary} \label{sec:summary}
We present multiwavelength, multi-epoch observations of five new CLQs and analyze the data both in the bright and faint states. Our main conclusions are as follows:

\begin{itemize}
\item Our optical spectroscopic follow-up confirms CLQ behavior, with continuum emission dramatically dimming, accompanied by fading of broad Balmer emission lines, such as H$\beta$, H$\gamma$, and H$\delta$. By selection, their broad H$\beta$ varies at $\gtrsim ~ 3\sigma$ level. The broad H$\alpha$ and \MgII\ emission varies less (or more slowly) than H$\beta$.

\item The continuum emission from CLQs varies across the electromagnetic spectrum, in X-ray, optical, and MIR. 

\item  The intrinsic X-ray continuum strength changes together with the optical continuum strength.
The X-ray power-law slope changes following a harder-when-fainter trend.  No strong absorption is detected in the faint-state X-ray spectra, so that a changing-obscuration model does not match the X-ray observations.

\item The large-amplitude MIR variability ($>0.5$ mag) detected in all five quasars is also inconsistent with the changing-obscuration model. 

\item The MIR variability for CLQs follows the variability in the optical, with a time lag consistent with the typical light-crossing time of the dusty torus for QSOs with robust lag measurements.

\item It is highly likely that the changes in these CLQs are due to changing accretion rate of the SMBH, so the multiwavelength emission varies accordingly. 

\item There are many similarities between the behavior of these CLQs and XRBs, indicating similarities in the BH accretion spanning factors of 10$^{8}$ in mass between stellar mass black holes and SMBH.
\end{itemize}

Only a handful of CLQ have been observed in X-rays both before and after the state transition. More such observations are needed to characterize their behavior more generally. Our five CLQs are basically at $\lambda_{\rm Edd} \gtrsim 10^{-2}$. More CLQ transitions over a wide range of $\lambda_{\rm Edd}$ will be important for studies of BH accretion processes and AGN structure. 

\begin{acknowledgments} 
We thank the referee for useful comments that improved the manuscript. We thank Ruancun Li and Malgorzata Sobolewska for useful discussions and suggestions. We thank Sebastian Gomez for help with observations and data reduction of the Magellan spectrum.

This research has made use of data obtained from the {\it Chandra} Data Archive and the {\it Chandra} Source Catalog, and software provided by the {\it Chandra} X-ray Center (CXC) in the application packages CIAO and Sherpa.  Q.Y. is partially supported for this work by the National Aeronautics and Space Administration through Chandra Award Numbers GO9-20086X and GO0-21084X, issued by the Chandra X-ray Center, which is operated by the Smithsonian Astrophysical Observatory for and on behalf of the National Aeronautics Space Administration under contract NAS8-03060. 

RMP acknowledges support from the National Science Foundation under grant No. 2206123.

This work used observations obtained with {\it XMM-Newton}.
Based on observations obtained with XMM-Newton, an ESA science mission with instruments and contributions directly funded by ESA Member States and NASA.

We acknowledge the use of telescopes, including MMT, Gemini, Magellan, the 3.5m telescope at APO, and the 1.2m telescope at the Fred Lawrence Whipple Observatory, operated by the Smithsonian Institution.
Observations reported here were obtained at the MMT Observatory, a joint facility of the Smithsonian Institution and the University of Arizona.
Based on observations obtained at the international Gemini Observatory, a program of NSF’s NOIRLab, which is managed by the Association of Universities for Research in Astronomy (AURA) under a cooperative agreement with the National Science Foundation on behalf of the Gemini Observatory partnership: the National Science Foundation (United States), National Research Council (Canada), Agencia Nacional de Investigaci\'{o}n y Desarrollo (Chile), Ministerio de Ciencia, Tecnolog\'{i}a e Innovaci\'{o}n (Argentina), Minist\'{e}rio da Ci\^{e}ncia, Tecnologia, Inova\c{c}\~{o}es e Comunica\c{c}\~{o}es (Brazil), and Korea Astronomy and Space Science Institute (Republic of Korea).
This paper includes data gathered with the 6.5 meter Magellan Telescopes located at Las Campanas Observatory, Chile.
The Hobby-Eberly Telescope Board of Directors approved the following update to the HET's Publication Policy in Fall 2023.  All publications that include HET data are expected to comply with the policy, which involves acknowledgements of the telescope and instrumentation, and appropriate citations of supporting publications.  When a peer-reviewed paper using HET data appears in print, the lead author should contact the HET Publications Coordinator, currently Donald Schneider (dps7@psu.edu), with the final journal reference information.
Based on observations obtained with the Apache Point Observatory 3.5-meter telescope, which is owned and operated by the Astrophysical Research Consortium.

We acknowledge the use of SDSS data. Funding for SDSS-III has been provided by the Alfred P. Sloan Foundation, the Participating Institutions, the National Science Foundation, and the U.S. Department of Energy Office of Science. The SDSS-III website is \url{http://www.sdss3.org/}. SDSS-III is managed by the Astrophysical Research Consortium for the Participating Institutions of the SDSS-III Collaboration including the University of Arizona, the Brazilian Participation Group, Brookhaven National Laboratory, Carnegie Mellon University, University of Florida, the French Participation Group, the German Participation Group, Harvard University, the Instituto de Astrofisica de Canarias, the Michigan State/Notre Dame/JINA Participation Group, Johns Hopkins University, Lawrence Berkeley National Laboratory, Max Planck Institute for Astrophysics, Max Planck Institute for Extraterrestrial Physics, New Mexico State University, New York University, Ohio State University, Pennsylvania State University, University of Portsmouth, Princeton University, the Spanish Participation Group, University of Tokyo, University of Utah, Vanderbilt University, University of Virginia, University of Washington, and Yale University.

This research has made use of PS1, WISE, CRTS, and PTF imaging data. The PS1 has been made possible through contributions by the Institute for Astronomy, the University of Hawaii, the Pan-STARRS Project Office, the Max-Planck Society and its participating institutes, the Max Planck Institute for Astronomy, Heidelberg and the Max Planck Institute for Extraterrestrial Physics, Garching, The Johns Hopkins University, Durham University, the University of Edinburgh, Queen's University Belfast, the Harvard-Smithsonian Center for Astrophysics, the Las Cumbres Observatory Global Telescope Network Incorporated, the National Central University of Taiwan, the Space Telescope Science Institute, the National Aeronautics and Space Administration under Grant No. NNX08AR22G issued through the Planetary Science Division of the NASA Science Mission Directorate, the National Science Foundation under Grant No. AST-1238877, the University of Maryland, and Eotvos Lorand University (ELTE). This publication makes use of data products from the \emph{Wide-field Infrared Survey Explorer}, which is a joint project of the University of California, Los Angeles, and the Jet Propulsion Laboratory/California Institute of Technology, funded by the National Aeronautics and Space Administration. The Catalina Sky Survey (CSS) is funded by the National Aeronautics and Space Administration under Grant No. NNG05GF22G issued through the Science Mission Directorate Near-Earth Objects Observations Program. The CRTS survey is supported by the US National Science Foundation under grants AST-0909182 and AST-1313422. We acknowledge the use of PTF data, and the website is \url{https://www.ptf.caltech.edu}.

We acknowledge the use of ZTF data. Based on observations obtained with the Samuel Oschin Telescope 48-inch and the 60-inch Telescope at the Palomar Observatory as part of the Zwicky Transient Facility project. ZTF is supported by the National Science Foundation under Grant No. AST-2034437 and a collaboration including Caltech, IPAC, the Weizmann Institute for Science, the Oskar Klein Center at
Stockholm University, the University of Maryland, Deutsches Elektronen-Synchrotron and Humboldt University, the TANGO
Consortium of Taiwan, the University of Wisconsin at Milwaukee, Trinity College Dublin, Lawrence Livermore National
Laboratories, and IN2P3, France. Operations are conducted by COO, IPAC, and UW.

\end{acknowledgments} 

\facilities{{\it Chandra}, {\it XMM-Newton}, MMT (Blue Channel spectrograph, BinoSpec), Magellan:Baade (IMACS), Gemini (GMOS-N), APO 3.5m (DIS), FLWO:1.2m, Sloan, PS1, PTF, CRTS, ZTF, WISE} 

\software{CIAO, SAS, XSPEC, CASA, IRAF, QSOFIT, X-CIGALE}

\appendix

\section{Extinction Coefficients} \label{sec:ext}

We provide a table of conversion coefficients from $E(B-V)_{\rm SFD}$ to extinction in 35 filters for the surveys we used in this work in Table \ref{tab:Extinction}. 
We calculated the extinction coefficients following the procedures in \citet{Schlafly2011}. The extinction in band $b$ is expressed as, 
\begin{equation}
A_b = -2.5 \log \left[ \frac{\int d\lambda W_b(\lambda) S(\lambda)10^{-A(\lambda)\Delta m_{1\mathrm{\mu m}}/2.5}}{\int d\lambda W(\lambda)S(\lambda)} \right]
\end{equation}
where $W_b(\lambda)$ is the filter throughput curve in band $b$ as a function of wavelength $\lambda$, $S$ is the source spectrum, and $A$ is the extinction law \citep{Fitzpatrick1999}, normalized to $A_{1 \mu {\rm m}} = 1$. $\Delta m_{1 \mu {\rm m}}$ is a normalization $N=0.78$ times the extinction at 1$\mu $m according to \citet{Schlegel1998}. The source spectrum S is a synthetic stellar spectrum, with $T_{\rm eff} = 7000~$K, log$~Z=-1$, and log$~g=4.5$ from \citet[][2500-10500$\AAm$]{Munari2005}, extrapolating into the infrared following a blackbody spectrum $S(\lambda) \propto \lambda^{-3}$. We compute the extinction coefficients $A_b / E(B-V)$, with $R_V =$ 2.1, 3.1, 4.1, and 5.1. 

\setcounter{table}{0}
\renewcommand{\thetable}{A\arabic{table}}

\begin{deluxetable}{cccccc}
\tablecaption{Extinction Coefficient \label{tab:Extinction}}
\tablewidth{1pt}
\tablehead{
\colhead{Survey} &
\colhead{Band} &
\colhead{} &
\colhead{$A_b/E(B-V)$} &
\colhead{}\\\cmidrule{3-6}
& & $R_V=2.1$ & $R_V=3.1$ & $R_V=4.1$ & $R_V=5.1$ 
}
\startdata
SDSS & $u$ & 5.443 & 4.247 & 3.716 & 3.418 \\
SDSS & $g$ & 3.852 & 3.304 & 3.052 & 2.907 \\
SDSS & $r$ & 2.260 & 2.287 & 2.298 & 2.305 \\
SDSS & $i$ & 1.585 & 1.697 & 1.749 & 1.780 \\
SDSS & $z$ & 1.211 & 1.263 & 1.286 & 1.300 \\
\hline
PS1 & $g$ & 3.666 & 3.188 & 2.967 & 2.840 \\
PS1 & $r$ & 2.254 & 2.278 & 2.288 & 2.294 \\
PS1 & $i$ & 1.569 & 1.682 & 1.733 & 1.763 \\
PS1 & $z$ & 1.259 & 1.323 & 1.352 & 1.369 \\
PS1 & $y$ & 1.087 & 1.104 & 1.111 & 1.116 \\
\hline
PTF & $g$ & 3.812 & 3.280 & 3.035 & 2.893 \\
PTF & $R$ & 2.002 & 2.087 & 2.126 & 2.148 \\
\hline
ZTF & $g$ & 3.748 & 3.240 & 3.005 & 2.870 \\
ZTF & $r$ & 2.111 & 2.169 & 2.195 & 2.210 \\
ZTF & $i$ & 1.444 & 1.542 & 1.587 & 1.612 \\
\hline
FLWO & $g$ & 3.758 & 3.245 & 3.008 & 2.871 \\
FLWO & $r$ & 2.226 & 2.254 & 2.266 & 2.273 \\
FLWO & $i$ & 1.524 & 1.631 & 1.680 & 1.708 \\
\hline
2MASS & $J$ & 0.778 & 0.722 & 0.696 & 0.681 \\
2MASS & $H$ & 0.517 & 0.457 & 0.430 & 0.414 \\
2MASS & $K_s$ & 0.344 & 0.308 & 0.292 & 0.282 \\
\hline
UKIRT & $Z$ & 1.226 & 1.282 & 1.308 & 1.323 \\
UKIRT & $Y$ & 0.989 & 0.978 & 0.973 & 0.970 \\
UKIRT & $J$ & 0.765 & 0.707 & 0.680 & 0.664 \\
UKIRT & $H$ & 0.524 & 0.464 & 0.437 & 0.421 \\
UKIRT & $K$ & 0.335 & 0.301 & 0.285 & 0.276 \\
\hline
VISTA & $Z$ & 1.247 & 1.307 & 1.335 & 1.352 \\
VISTA & $Y$ & 1.003 & 0.996 & 0.993 & 0.991 \\
VISTA & $J$ & 0.763 & 0.705 & 0.678 & 0.662 \\
VISTA & $H$ & 0.520 & 0.461 & 0.433 & 0.417 \\
VISTA & $K_s$ & 0.351 & 0.314 & 0.298 & 0.288 \\
\hline
WISE & $W1$ & 0.180 & 0.176 & 0.174 & 0.173 \\
WISE & $W2$ & 0.112 & 0.120 & 0.123 & 0.125 \\
WISE & $W3$ & 0.034 & 0.046 & 0.052 & 0.055 \\
WISE & $W4$ & 0.014 & 0.022 & 0.026 & 0.028 \\
\enddata
\end{deluxetable}

\section{SED Data}
We compiled the multiwavelength photometric data in both states (summarized in Table \ref{tab:SED_data}). Since we simultanously fit the X-ray to MIR SED, we chose the photometric data closest in time to the X-ray epochs (indicated by the ``Epoch" column in Table \ref{tab:Xray_obs}). 

\setcounter{table}{0}
\renewcommand{\thetable}{B\arabic{table}}

\begin{deluxetable*}{cccclcc}
\tabletypesize{\tiny}
\tablecaption{Photometric Data for CLQ SED \label{tab:SED_data}}
\tablewidth{1pt}
\tablehead{
\colhead{Name} &
\colhead{Epoch} &
\colhead{Survey/Telescope} &
\colhead{Band} &
\colhead{MJD} &
\colhead{mag} &
\colhead{uncertainty}
}
\startdata
J0206 & 1 & PS1 & $g$ & 56631.333  & 20.125  & 0.040 \\
J0206 & 1 & PS1 & $r$ & 56988.307  & 20.209  & 0.062 \\
J0206 & 1 & PS1 & $i$ & 56916.486  & 19.685  & 0.038 \\
J0206 & 1 & PS1 & $z$ & 56644.232  & 19.047  & 0.039 \\
J0206 & 1 & PS1 & $y$ & 56517.627  & 18.521  & 0.037 \\
J0206 & 1 & WISE & $W1$ & 56860.281 & 14.762 & 0.019\\
J0206 & 1 & WISE & $W2$ & 56860.281 & 14.046 & 0.041\\
J0206 & 1 & WISE & $W3$ & 55312.068 & 11.487 & 0.134\\
J0206 & 1 & WISE & $W4$ & 55312.068 & 8.877 & 0.775\\
\hline
J0206 & 2 & FLWO & $g$ & 58481.133  & 20.816  & 0.100 \\
J0206 & 2 & FLWO & $r$ & 58465.266  & 19.815  & 0.046 \\
J0206 & 2 & WISE & $W1$ & 58480.006 & 15.748 & 0.051\\
J0206 & 2 & WISE & $W2$ & 58480.006 & 14.855 & 0.092\\
\hline
J0224 & 1 & PS1 & $g$ & 55854.412  & 18.906  & 0.019 \\
J0224 & 1 & PS1 & $r$ & 55854.462  & 18.655  & 0.015 \\
J0224 & 1 & PS1 & $i$ & 56209.550  & 18.489  & 0.013 \\
J0224 & 1 & PS1 & $z$ & 55583.258  & 18.118  & 0.017 \\
J0224 & 1 & PS1 & $y$ & 55584.260  & 18.035  & 0.050 \\
J0224 & 1 & VHS & $J$ & 56298.146  & 16.793  & 0.009 \\
J0224 & 1 & VHS & $K_s$ & 56298.136  & 15.108  & 0.011 \\
J0224 & 1 & WISE & $W1$ & 55213.406 & 13.599 & 0.009\\
J0224 & 1 & WISE & $W2$ & 55213.406 & 12.771 & 0.014\\
J0224 & 1 & WISE & $W3$ & 55318.732 & 10.551 & 0.052\\
J0224 & 1 & WISE & $W4$ & 55322.150 & 7.589 & 0.220\\
\hline
J0224 & 2 & ZTF & $g$ & 59138.359  & 20.058  & 0.025 \\
J0224 & 2 & ZTF & $r$ & 59158.345  & 18.976  & 0.009 \\
J0224 & 2 & WISE & $W1$ & 59054.290 & 14.244 & 0.013\\
J0224 & 2 & WISE & $W2$ & 59054.290 & 13.546 & 0.030\\
\hline
J0829 & 1 & PTF & $r$ & 56724.155  & 20.166  & 0.119 \\
J0829 & 1 & UHS & $J$ & 56066-57784 & 18.356  & 0.073 \\
J0829 & 1 & WISE & $W1$ & 56764.247 & 15.347 & 0.114\\
J0829 & 1 & WISE & $W2$ & 56764.247 & 14.168 & 0.153\\
J0829 & 1 & WISE & $W3$ & 55299.743 & 11.789 & 0.228\\
J0829 & 1 & WISE & $W4$ & 55299.743 & 9.536 & 0.929\\
\hline
J0829 & 2 & ZTF & $g$ & 58891.198  & 21.285  & 0.036 \\
J0829 & 2 & ZTF & $r$ & 58891.290  & 20.826  & 0.025 \\
J0829 & 2 & ZTF & $i$ & 58257.172  & 19.485  & 0.108 \\
J0829 & 2 & WISE & $W1$ & 58778.937 & 16.046 & 0.058\\
J0829 & 2 & WISE & $W2$ & 58778.937 & 15.322 & 0.138\\
\hline
J1226 & 1 & SDSS & $u$ & 51923  & 18.936  & 0.042 \\
J1226 & 1 & SDSS & $g$ & 51923  & 18.699  & 0.021 \\
J1226 & 1 & SDSS & $r$ & 51923  & 19.262  & 0.036 \\
J1226 & 1 & SDSS & $i$ & 51923  & 18.822  & 0.027 \\
J1226 & 1 & SDSS & $z$ & 51923  & 19.078  & 0.094 \\
J1226 & 1 & UKIDSS & $Y$ & 54398  & 18.387  & 0.047 \\
J1226 & 1 & UKIDSS & $J$ & 54398  & 17.772  & 0.054 \\
J1226 & 1 & UKIDSS & $H$ & 54398  & 17.135  & 0.043 \\
J1226 & 1 & UKIDSS & $K_s$ & 54398  & 16.148  & 0.034 \\
J1226 & 1 & WISE & $W1$ & 55370.140 & 14.304 & 0.013\\
J1226 & 1 & WISE & $W2$ & 55370.140 & 13.302 & 0.020\\
J1226 & 1 & WISE & $W3$ & 55370.743 & 10.519 & 0.073\\
J1226 & 1 & WISE & $W4$ & 55370.743 & 8.221 & 0.565\\
\hline
J1226 & 2 & ZTF & $g$ & 58607.299  & 20.751  & 0.061 \\
J1226 & 2 & ZTF & $r$ & 58583.255  & 20.674  & 0.052 \\
J1226 & 2 & ZTF & $i$ & 58261.189  & 19.846  & 0.097 \\
J1226 & 2 & WISE & $W1$ & 58638.071 & 15.005 & 0.024\\
J1226 & 2 & WISE & $W2$ & 58638.071 & 14.315 & 0.058\\
\hline
J1338 & 1 & SDSS & $u$ & 52053  & 19.884  & 0.041 \\
J1338 & 1 & SDSS & $g$ & 52053  & 19.437  & 0.025 \\
J1338 & 1 & SDSS & $r$ & 52053  & 19.032  & 0.020 \\
J1338 & 1 & SDSS & $i$ & 52053  & 18.734  & 0.018 \\
J1338 & 1 & SDSS & $z$ & 52053  & 18.433  & 0.036 \\
J1338 & 1 & UKIDSS & $Y$ & 54989  & 17.700  & 0.026 \\
J1338 & 1 & UKIDSS & $J$ & 54989  & 17.154  & 0.029 \\
J1338 & 1 & UKIDSS & $H$ & 54989  & 16.437  & 0.017 \\
J1338 & 1 & UKIDSS & $K_s$ & 54989  & 15.515  & 0.014 \\
J1338 & 1 & WISE & $W1$ & 55211.722 & 13.818 & 0.010\\
J1338 & 1 & WISE & $W2$ & 55211.722 & 13.170 & 0.020\\
J1338 & 1 & WISE & $W3$ & 55315.574 & 10.741 & 0.069\\
J1338 & 1 & WISE & $W4$ & 55387.109 & 7.400 & 0.354\\
\hline
J1338 & 2 & FLWO & $g$ & 57887.197  & 19.578  & 0.032 \\
J1338 & 2 & FLWO & $r$ & 57887.201  & 19.008  & 0.026 \\
J1338 & 2 & WISE & $W1$ & 57765.450 & 15.033 & 0.029\\
J1338 & 2 & WISE & $W2$ & 57765.450 & 14.244 & 0.060\\
\enddata
\tablecomments{
We use photometry in optical, near-infrared, and mid-infrared that is closest to the X-ray observation epochs.
We use PSF (AB) magnitudes in the optical, \SI{2}{\arcsecond} diameter aperture (Vega) magnitudes in the near-infrared. The WISE data is in Vega magnitude.}
\end{deluxetable*}

\bibliography{references_CXO.bib}

\end{document}